\documentclass[aps,pre,twocolumn,groupedaddress,showpacs,superscriptaddress,amssymb,amsmath]{revtex4-1}
\usepackage[utf8]{inputenc}
\usepackage{mathtools}
\usepackage{graphicx}
\usepackage{tabularx}
\usepackage{xcolor}
\usepackage{amsmath}
\DeclareMathOperator{\sech}{sech}

\usepackage{comment}
\newcommand{\be}{\begin{equation}}
\newcommand{\ee}{\end{equation}}
\newcommand{\bea}{\begin{eqnarray}}
\newcommand{\eea}{\end{eqnarray}}

\usepackage{dcolumn}
\usepackage{hyperref}
\hypersetup{
    colorlinks=true,
    linkcolor=blue,
	citecolor=cyan
}
\usepackage{bm}
\usepackage{epsf}
\usepackage{subfigure}
\usepackage{braket}
\usepackage{tensor}
\usepackage{soul}
\usepackage{float}
\usepackage{dblfloatfix} 
\usepackage{lipsum}

\usepackage{mathrsfs}

\usepackage{amsfonts}

\makeatletter
\newsavebox{\@brx}
\newcommand{\llangle}[1][]{\savebox{\@brx}{\(\m@th{#1\langle}\)}%
  \mathopen{\copy\@brx\kern-0.5\wd\@brx\usebox{\@brx}}}
\newcommand{\rrangle}[1][]{\savebox{\@brx}{\(\m@th{#1\rangle}\)}%
  \mathclose{\copy\@brx\kern-0.5\wd\@brx\usebox{\@brx}}}
\makeatother

\DeclareMathOperator{\trace}{Tr}
\def\lm{\lambda}
\def\la{\langle}
\def\ra{\rangle}
\def\l{\Big(}
\def\r{\Big)}
\def\om{{\omega}}
\def\tcs{\tanh^2{\Big(\frac{\beta_c\omega_0}{2}\Big)}}
\def\ths{\tanh^2{\Big(\frac{\beta_h\omega_1}{2}\Big)}}
\def\tc{\tanh{\Big(\frac{\beta_c\omega_0}{2}\Big)}}
\def\th{\tanh{\Big(\frac{\beta_h\omega_1}{2}\Big)}}

\def\ccs{\coth^2{\Big(\frac{\beta_c\omega_0}{2}\Big)}}
\def\chs{\coth^2{\Big(\frac{\beta_h\omega_1}{2}\Big)}}
\def\cc{\coth{\Big(\frac{\beta_c\omega_0}{2}\Big)}}
\def\ch{\coth{\Big(\frac{\beta_h\omega_1}{2}\Big)}}

\date{\today}

\begin{document}
\title{Study of bounds on non-equilibrium fluctuations for asymmetrically driven quantum Otto engines}

\author{Sandipan Mohanta}
\email{mohanta.sandipan@students.iiserpune.ac.in}
\affiliation{Department of Physics, Indian Institute of Science Education and Research, Pune 411008, India}
\author{Madhumita Saha}
\email{madhumita.saha@acads.iiserpune.ac.in} 
\affiliation{Department of Physics,
		Indian Institute of Science Education and Research, Pune 411008, India}
\author{B. Prasanna  Venkatesh}
\email{prasanna.b@iitgn.ac.in} 
\affiliation{Department of Physics, Indian Institute of Technology Gandhinagar, Palaj, Gujarat 382355, India}

\author {Bijay Kumar Agarwalla}
\email{bijay@iiserpune.ac.in}
\affiliation{Department of Physics, Indian Institute of Science Education and Research, Pune 411008, India}

\date{\today}
\begin{abstract}
For a four-stroke asymmetrically driven quantum Otto engine with working medium modeled by a single qubit, we study the bounds on non-equilibrium fluctuations of work and heat. We find strict relations between the fluctuations of work and individual heat for hot and cold reservoirs in arbitrary operational regimes. Focusing on the engine regime, we show that the ratio of non-equilibrium fluctuations of output work to input heat from the hot reservoir is both upper and lower bounded. As a consequence, we establish hierarchical relation between the relative fluctuations of work and heat for both cold and hot reservoirs and further make a connection with the thermodynamic uncertainty relations. We discuss the fate of these bounds also in the refrigerator regime. The reported bounds, for such asymmetrically driven engines, emerge once both the time-forward and the corresponding {reverse} cycles of the engine are considered on an equal footing. We also extend our study and report bounds for a parametrically driven harmonic oscillator Otto engine.
\end{abstract}

\maketitle
\section{Introduction}
Thermodynamic devices have significantly advanced humankind since the Industrial Revolution in 1760, whether it be through the development of steam engines in the past or the modern automobiles we use every day. {Improving these machines' performance has long been a top concern.} The two thermodynamic machine types that are most frequently utilised are (i) heat engines and (ii) refrigerators. While the latter uses external work to remove heat from a cold body, the former's goal is to transform heat absorbed from a hot body into work \cite{Callen-book,Carnot-2}. 

Sadi Carnot's foundational research established a general upper bound on the efficiency of any heat engine operating between hot and cold reservoirs at constant inverse temperatures, $\beta_h$ and $\beta_c$, respectively. This upper bound, commonly known as the Carnot bound \cite{Callen-book, Carnot-2}, is stated as $\eta_c=1\!-\!\beta_h/\beta_c$. 
{Similarly, for any refrigerator operating between the same two reservoirs, a maximum achievable coefficient of performance (COP), also referred to as the cooling efficiency, $\varepsilon_c=(1\!-\!\eta_c)/\eta_c$, was reported}   \cite{Callen-book, otto-refrigerator}.
 
These bounds, however, can be saturated only in an ideal reversible thermodynamic cycle, which requires infinite cycle time and thus have very limited practical importance. Understanding the finite-time behaviour of thermodynamic cycles, where inevitable irreversibility may lead to tighter bounds on efficiency, is thus a more practical but demanding task \cite{engine-expt-spin,otto-ho-mdpi,Otto-friction-effect,Otto-09,Wang_2007,Chand_2021,Gernot-collective_performance_Otto,cangemi2023quantum}. 
Another important point to note is that, with massive technological advancements, miniaturisation of thermodynamic devices is now possible, and enormous efforts have been invested in studying thermodynamics of such small-scale systems \cite{engine-expt-spin,Eric-engine,engine-refg,ion-heat-engine,Zhang2022,blickle2012realization,martinez2016brownian,martinez2017colloidal,myers2022quantum,Amit_Dutta-review}, leading to the emergence of new disciplines such as stochastic thermodynamics and quantum thermodynamics \cite{Seifert2012,Seifert2008, Vinjanampathy2016}.

The performance of small-scale thermodynamic devices can be significantly impacted by fluctuations of both thermal and quantum origin, making their study crucial. It is a well established fact that fluctuations in equilibrium systems are linked to response functions by the celebrated fluctuation-dissipation theorem \cite{kubo1966fluctuation, FD1}. Over the last few decades, via the discovery of the fluctuation relations \cite{fluc-1,fluc-2,fluc-3}, great progress has also been made towards understanding non-equilibrium fluctuations. Very recently, understanding the effects of such non-equilibrium fluctuations and the higher-order statistical moments on the efficiency of small-scale engines, whether classical or quantum, has attracted a lot of attention \cite{holubec2021fluctuations, watanabe2022finite, Gentaro-1, Saryal2021,Universal-Agarwalla}. In this context, it was recently shown that the ratio of output work fluctuation to input heat fluctuation is upper bounded by the square of the Carnot bound in the engine operational regime \cite{Gentaro-1}. Another work demonstrated that the square of mean efficiency of a symmetrically driven qubit-Otto  engine sets a lower bound on the ratio of output work fluctuation to input heat fluctuation. \cite{Saryal2021}.

Motivated by such works, in this paper, we investigate a generalised asymmetrically driven four-stroke quantum Otto-cycle \cite{otto-ho-mdpi,engine-expt-spin,Otto-01,Otto-02,Otto-03,Otto-04,Otto-05,Otto-06,Otto-07,Otto-08,Otto-09,Otto-10,Lutz1,Zheng2016, Potts_TUR, Horowitz_TUR_2019,Jaramillo_2016,Victor_2021,Chand_2017,Shastri2022,Campisi-PRB-2020}. We concentrate on the paradigmatic qubit-Otto cycle, which was recently implemented in the NMR platform \cite{engine-expt-spin}, and we offer fundamental insights into the relationships between non-equilibrium fluctuations of work and heat from both hot and cold reservoirs while operating in various operational regimes. Furthermore, we provide hierarchical relationships between the relative fluctuations in the engine operational regime and connect our results with the thermodynamic uncertainty relations (TURs) \cite{Barato:2015:UncRel,trade-off-engine,Gingrich:2016:TUR,Falasco,Sacchi-TUR-HO,Landi-Goold-TUR,multiterminal-TUR,Bijay-Dvira-TUR}. We also explore similar constraints on fluctuations for a harmonic oscillator working medium \cite{otto-ho-mdpi,otto-ho-squeezed-bath,ion-heat-engine}.

We organize our paper as follows: In Sec.~\ref{sec:Quantum-Otto}, we give a brief introduction to the four-stroke quantum Otto cycle that is asymmetrically driven and and also summarize the two-time measurement scheme to obtain the joint heat and work statistics. In Sec.~\ref{sec:s-qubit}, we initially consider a single qubit  as the working medium and derive the central results of the paper, namely, the demonstration of useful bounds on non-equilibrium fluctuations of heat and work, independent of its regime of operation. Following this, we obtain the bounds for the engine regime of operation as well as make connections of the derived bounds with the TURs. In addition, we also discuss the bounds for the refrigerator regime of operation and supplement our analytical calculations with numerical results for  better illustration. In Sec.~\ref{sec:HO} we discuss briefly the results for the parametrically driven harmonic oscillator working medium. Finally we summarize our central results in Sec.~\ref{sec:summary}. Certain details of the calculations are provided in the appendices.

\section{Quantum Otto cycle and Two-point measurement scheme}
\label{sec:Quantum-Otto}
\begin{figure}[H]
    \centering
     \includegraphics[width=\columnwidth]{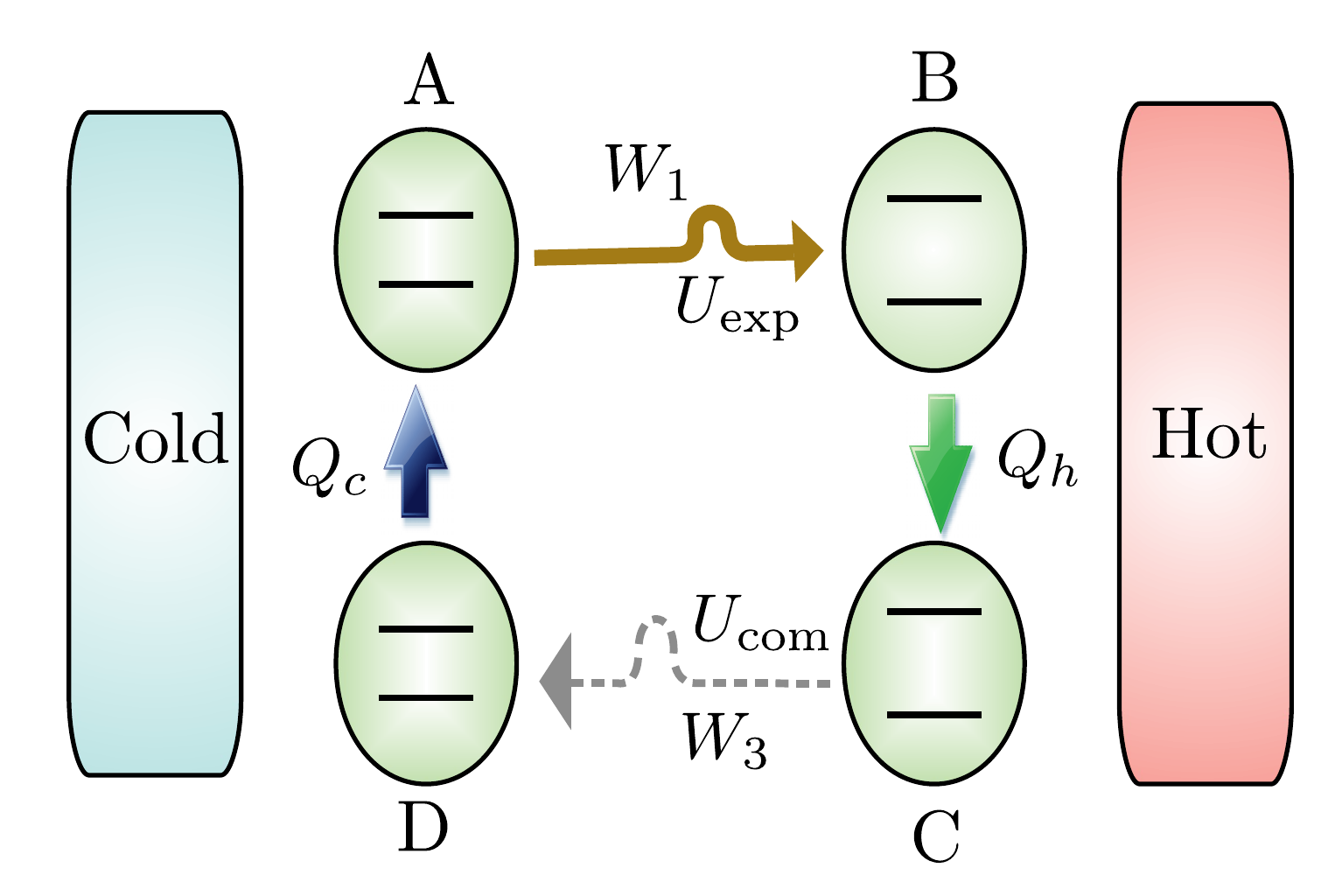}
     \caption{Schematic for a finite-time asymmetrically driven quantum Otto cycle: Two unitary strokes $(A \!\to\! B)$ and $(C \!\to\! D)$ correspond to expansion and compression of the working medium and governed by $U_\mathrm{exp}$ and $U_\mathrm{com}$, respectively. For symmetric driving,  $U_{\rm com} \!=\! \Theta{U_{\rm exp}^\dagger}\Theta^\dagger$. The thermalization strokes are represented by $(B \!\to\! C)$ for the hot end and $(D\! \to\! A)$ for the cold end.} 
     \label{fig:1}
\end{figure}
First, we briefly demonstrate the various strokes of an Otto cycle (see Fig.~\ref{fig:1}). {The working medium is characterized by the parametric time-dependent Hamiltonian $H[\lambda_t]$, and the system parameter $\lm_t$ is varied between $\lm_i$ and $\lm_f$ during one cycle. The system begins at $t\!=\!0$ with Hamiltonian $H[\lm_i]$ in a thermal state with an inverse temperature $\beta_c$ and passes through the following four steps:} 

(i) {\it Expansion stroke} $(A\!\to\! B)$: In the first step, system eigen-energy spacing is enlarged by external drive as the system Hamiltonian changes unitarily from {$H[\lambda_i]$ to $H[\lambda_f]$ in $\tau_1$ time duration}. This stroke is dictated by a unitary generator {$U_\mathrm{exp}$}, and the system consumes stochastic work $W_1$ during this stroke. Note that, the volume compression stroke in a traditional Otto cycle is comparable to this stroke.

(ii){\it Heating stroke} ($B\!\to\! C$): In the second step, the system is brought into weak contact with the hot reservoir at inverse temperature $\beta_h$. With no changes to the external parameters, the system Hamiltonian remains constant, and the system absorbs heat $Q_h$ from the hot reservoir during the time duration $\tau_2$ of this stroke and thermalizes.

(iii){\it Compression stroke} ($C\!\to \!D$): The system is separated from the hot reservoir during the third step, and an external driving source causes the system Hamiltonian to transition from {$H[\lambda_f]$ back to $H[\lambda_i]$} in stroke time duration $\tau_3$. {$U_\mathrm{com}$} is the unitary generator of this stroke. The system's energy gap narrows during this stroke, producing stochastic work $W_3$. It is comparable to the  volume expansion stroke of the traditional cycle.

(iv){\it Cooling stroke} ($D\!\to\! A$): In the final step, the system is brought back into weak contact with the cold reservoir, keeping the system Hamiltonian constant, to complete the cycle. The system thermalizes during this stroke in time $\tau_4$, and heat $Q_c$ is exchanged.


Throughout our discussion, we adhere to the sign convention that energy entering the system is positive. In order to guarantee complete thermalization of the working medium during the heat exchange strokes, we further assume that the durations of these strokes, $\tau_2$ and $\tau_4$, are much longer than the system relaxation time.
We emphasise that, the compression stroke mentioned here is not carried out by time-reversing the expansion technique, \emph{i.e.}, { $U_\mathrm{com}\ne \Theta U_\mathrm{exp}^\dagger\Theta^\dagger$, where $\Theta$ is the  anti-unitary time reversal operator.} This breaks the underlying time-reversal symmetry connecting the expansion and compression strokes, making the time forward and {reverse} cycles distinct.
This leads to differing joint probability distributions of stochastic variables-- work and heat. 
In this study, we also make the assumption that the energy needed to couple and decouple the system from the reservoirs, before and after the heat exchange strokes, are negligibly small in comparison to all other energy scales involved in the problem.


To construct the joint probability distribution (PD) of total work output $W\!=\!W_1+W_3$ and heat input $Q_h$ from the hot bath, we perform projective energy measurements \cite{fluc-1,fluc-2,fluc-3} on the respective Hamiltonians involved in the first three strokes $(A\!\to\! B\!\to\! C\!\to\! D)$. We write the joint PD  as,{
\begin{equation}
\label{PD}
\begin{aligned}
     P&(W,Q_h)\!=\!\sum_{n,m,k,l} \delta \Big(W\!-\!\epsilon_m[\lambda_f]\! +\! \epsilon_n[\lambda_i]\!-\!\epsilon_l[\lambda_i]\!+\!\epsilon_k[\lambda_f]\Big)\\&\delta\Big(Q_h\!-\!\epsilon_k[\lambda_f]\!+\!\epsilon_m[\lambda_f]\Big){\cal T}^{{I}}_{n\to m}{\cal T}^{II}_{k\to l} 
              \frac{e^{-\beta_c \epsilon_n[\lambda_i]}}{{\cal Z}_c[\lambda_i]}
              \frac{e^{-\beta_h \epsilon_k[\lambda_f]}}{{\cal Z}_h[\lambda_f]},
\end{aligned}              
\end{equation}
where, ${\cal Z}_{c}[\lambda_i]\!=\!\sum_n\exp(-\beta_c \epsilon_n[\lambda_i]),$ and ${\cal Z}_{h}[\lambda_f]\!=\!\sum_k\exp(-\beta_h \epsilon_k[\lambda_f]),$ are the canonical partition functions} and the unitarily driven transition probabilities are given as,{
\begin{gather}
   {\cal T}^{I}_{n \to m}=\big|\la m ;\lambda_f|U_\mathrm{exp}|n;\lambda_i\ra\big|^2,\\
   {\cal T}^{II}_{k \to l}=\big|\la l ;\lambda_i|U_\mathrm{com}|k;\lambda_f\ra\big|^2\label{transition-probability}.
\end{gather}
}Despite the fact that the PD is constructed here by measuring energy exchanges in three of the four strokes, we emphasize that, in case of perfect thermalization in the heat exchange strokes, the statistics of the energy exchange variable $Q_c$ in the fourth stroke can be easily obtained from the above distribution {(see Appendix \ref{sec:AppA} for details)}.

To obtain the averages and fluctuations (second cumulants) of work and heat we  study the  characteristic function (CF) of total work and heat defined by the Fourier transformation of PD,
\begin{align}\label{CF}
    \chi(\gamma_w,\gamma_h)=\int dW\,dQ_h\,P(W,Q_h)\,e^{i\gamma_w W}e^{i\gamma_h Q_h},
\end{align}
where $\gamma_w$ and $\gamma_h$ are the conjugate Fourier parameters for $W$ and $Q_h$, respectively. All the cumulants (denoted by double angular brackets) are easily obtained from this CF in Eq.~\eqref{CF} by taking partial derivatives with respect to $\gamma_w$ and $\gamma_h$,
\begin{align}
    \llangle W^n Q_h^m \rrangle = \frac{\partial^n\partial^m}{\partial(i\gamma_w)^n\partial(i\gamma_h)^m}\,\mathrm{ln}{\chi(\gamma_w,\gamma_h)}\Big|_{\gamma_w,\gamma_h = 0}.
\end{align}

It is crucial to emphasise that in order to derive bounds on non-equilibrium fluctuations for this asymmetric driving situation, {$U_\mathrm{com}\ne \Theta U_\mathrm{exp}^\dagger\Theta^\dagger$}, we {have taken} into account both the forward and {reverse} processes. The PD specified above will now be referred to as the forward PD, indicated by $P_{F}(W,Q_h)$. In a similar way, we construct the PD for the {reverse} cycle by following the strokes  $(D\!\to\! C\!\to\! B\!\to\! A)$. In the {reverse} cycle the expansion and compression strokes are accomplished by { $\tilde{U}_\mathrm{exp}\!=\!\Theta U_\mathrm{com}^\dagger\Theta^\dagger$ and $\tilde{U}_\mathrm{com}\!=\!\Theta U_\mathrm{exp}^\dagger\Theta^\dagger$}, respectively, and  the corresponding PD is denoted as $P_{R}(W,Q_h)$. This definition of the PDs for forward and {reverse} cycles is consistent with the fluctuation symmetry (see Appendix \ref{sec:AppB} for details),
\begin{align}\label{Fluctuation-relation}
    \frac{P_{F}(W,Q_h)}{P_{R}(-W,-Q_h)}=\exp{(\Sigma)},
\end{align}
where 
\begin{align}
\Sigma= \beta_c W+(\beta_c\!-\!\beta_h) Q_h \label{eq:StochEntropy}
\end{align}
is the total stochastic entropy production in the forward cycle. The fluctuation relation, expressed in terms of the CF, is $\chi_F\big(\gamma_w,\gamma_h\big)=\chi_R\big(\!-\!\gamma_w+i\beta_c,-\gamma_h+i(\beta_c\!-\!\beta_h)\big)$. { Note that this detailed fluctuation relation, also known as the heat engine fluctuation relation, was previously introduced in \cite{Campisi_2014} and holds true for   asymmetrically driven quantum Otto engine scenario considered here.}

With a clear understanding of the forward and {reverse} PDs, we now  define the {symmetrized} fluctuations and relative fluctuations (RF) of work and heat, taking into account both forward and {reverse} processes on an equal footing,
\begin{gather}
\Delta \phi\coloneqq{\llangle \phi^2\rrangle _F+\llangle \phi^2 \rrangle_R},\label{def-fluc}\\
\mathrm{RF}(\phi)\coloneqq2\frac {\llangle \phi^2\rrangle _F+\llangle \phi^2 \rrangle_R}{\big(\langle \phi \rangle_F+\langle \phi \rangle_R\big)^2},
\label{def-re}
\end{gather} 
where $\phi= W, Q_h, Q_c$ and $\langle .\rangle_{F(R)}$ and $\llangle .\rrangle_{F(R)}$ {denote} the first and second cumulants corresponding to the forward ({reverse}) PDs. Note that with these generalised definitions, the time-reversal symmetric case, with $P_{F}(W,Q_h)=P_{R}(W,Q_h)$, now becomes a specific case of our generalised method.
In what follows, we first start with our model-- asymmetrically driven single qubit as working medium and discuss bounds on non-equilibrium fluctuations for heat and work.

\section{Single qubit  under asymmetric driving as a working medium:}
\label{sec:s-qubit}
We consider a single qubit as the working medium of an Otto cycle. During the unitary expansion stroke $A\!\to\! B$, {the system Hamiltonian changes from $H[\om_0]$ with energy level spacing $\om_0$ to $H[\om_1]$ with enhanced level spacing $\om_1$ in time $\tau_1$ and returns to $H[\om_0]$ during the compression stroke $(C\!\to\! D)$ in time $\tau_3$.} The quasistatic limit cycle is reached when the unitary stokes are driven quantum-adiabatically, meaning no transition is allowed between the instantaneous energy eigen states {$\{|0;\om_t\rangle,|1;\om_t\rangle\}$.} From this point forward, we'll frequently use the words ``quasistatic'' and ``quantum-adiabatic'' interchangeably. Any departure from the ideal quasistatic cycle can therefore be quantified by the transition probabilities between the states. We  define, for the expansion stroke, {$u\!=\!|{\langle} 0;\om_1 | U_\mathrm{exp}|0;\om_0\rangle|^2\!=\!|{\langle}1;\om_1 | U_\mathrm{exp}|1;\om_0\rangle|^2$, as the probability of no transition between the system energy states and similarly for the compression stroke $v\!=\!|{\langle} 0;\om_0 | U_\mathrm{com}|0;\om_1\rangle|^2\!=\!|{\langle} 1;\om_0 | U_\mathrm{com}|1;\om_1\rangle|^2$.} Note that $u,v\in[0,1]$ and $u\!=\!v$  corresponds to the time-reversal symmetric situation. The quasistatic limit corresponds to $u\!=\!v\!=\!1$.
The exact expression of the forward CF for the asymmetric driving case can be obtained in a compact form and is given by
\begin{widetext}
\begin{align}\label{CF-Qubit}
     {\chi}_F(\gamma_w,\gamma_h)&= \l u\cos{\Big[\frac{1}{2}\big(\gamma_w(\omega_0\!-\!\omega_1)+\gamma_h\omega_1-i\beta_c\omega_0\big)\Big]}+(1\!-\!u)\cos{\Big[\frac{1}{2}\big(\gamma_w(\omega_0\!+\!\omega_1)-\gamma_h\omega_1-i\beta_c\omega_0\big)\Big]}\r\nonumber\\
     &\times\l v\cos{\Big[\frac{1}{2}\big(\gamma_w(\omega_0\!-\!\omega_1)+\gamma_h\omega_1+i\beta_h\omega_1\big)\Big]}+(1\!-\!v)\cos{\Big[\frac{1}{2}\big(-\gamma_w(\omega_0\!+\!\omega_1)+\gamma_h\omega_1+i\beta_h\omega_1\big)\Big]}\r\nonumber\\
     &\times\sech{\l\frac{\beta_c\omega_0}{2}\r}\sech{\l\frac{\beta_h\omega_1}{2}\r}.
 \end{align}
\end{widetext}
For the {reverse} cycle the corresponding {reverse} CF, denoted as $\chi_R(\gamma_w,\gamma_h)$, is simply obtained by the replacement $u\leftrightarrow v$ in Eq.~(\ref{CF-Qubit}).
\subsection{Expressions for averages and fluctuations of heat and work}
With the CF [Eq.~({\ref{CF-Qubit}})] in hand, one can write down the expressions of the first and second cumulants of work and heat as,
\begin{widetext}
\begin{gather}
     \langle Q_h \rangle_F = -\frac{\omega_1}{2} \Big[ \tanh{\left(\frac{\beta_h\omega_1}{2}\right)}+(1\!-\!2u)\tanh{\left(\frac{\beta_c\omega_0}{2}\right)} \Big],  \\
      \llangle Q_h^2\rrangle_F=\frac{\omega_1^2}{4}\Big[2-\tanh^2{\left(\frac{\beta_h\omega_1}{2}\right)}-(1\!-\!2u)^2 \tanh^2{\left(\frac{\beta_c\omega_0}{2}\right)}\Big], \\
      \langle W \rangle_F  = \langle W_1 \rangle_F+\langle W_3 \rangle_F =\frac{1}{2} \big( \omega_0 +(1\!-\!2u)\omega_1\big)\tanh{\left(\frac{\beta_c\omega_0}{2}\right)}+\frac{1}{2}\big(\omega_1+(1\!-\!2v)\omega_0\big)\tanh{\left(\frac{\beta_h\omega_1}{2}\right), \label{qubit-work}}\\
      \llangle W^2\rrangle_F=\frac{(\omega_0+\omega_1)^2}{2}-(u\!+\!v)\,\omega_0 \, \omega_1-\langle W_1\rangle^2_F-\langle W_3\rangle_F^2, \\
      \langle Q_c \rangle_F = -\frac{\omega_0 }{2}\Big[ \tanh{\left(\frac{\beta_c\omega_0}{2}\right)}+(1\!-\!2v)\tanh{\left(\frac{\beta_h\omega_1}{2}\right)} \Big],  \\
      \llangle Q_c^2\rrangle_F=\frac{\omega_0^2}{4}\Big[2-\tanh^2{\left(\frac{\beta_c\omega_0}{2}\right)}-(1\!-\!2v)^2 \tanh^2{\left(\frac{\beta_h\omega_1}{2}\right)}\Big] .
\end{gather}
\end{widetext}
In a similar way, the averages and fluctuations (second cumulants) for the {reverse} cycle can be obtained by interchanging $u$ and $v$. {Notice that, depending on values of $u$ and $v$,  the Otto-cycle (forward or reverse) has four possible operational regimes-- engine, refrigerator, heater, and accelerator \cite{Campisi-PRB-2020}. } 
\subsection{General lower bound on the ratio of work to heat fluctuations}
\label{sec:general_bound}

We now discuss the first  central result of our work where we demonstrate bounds on the non-equilibrium fluctuations of heat and work for a qubit system as the working fuel of an quantum Otto cycle. In particular we find the following lower bounds regardless of the operational regime of the Otto cycle:
 \begin{align}
    \frac{\Delta W}{\Delta Q_h}&\ge\Big(\frac{\om_1\!-\!\om_0}{\om_1}\Big)^2,\label{lbound}\\
    \frac{\Delta W}{\Delta Q_c}&>\Big(\frac{\om_1\!-\!\om_0}{\om_1}\Big)^2.\label{lboundc}
\end{align}
 Recall that, as stated in Eq.~(\ref{def-fluc}), the definition of fluctuations include both forward and {reverse} processes. We now provide a proof for these results.


Let us first consider the fluctuations of heat from the hot reservoir. Our strategy will be to consider a modified version of Eq.~\eqref{lbound} written as $\om_1^2\Delta W\!-\!(\om_1\!-\!\om_0)^2\Delta Q_h \geq 0$. With a bit of straightforward calculations, the left hand side of this inequality can be written as
    \begin{align}
        &\om_1^2\Delta W-(\om_1\!-\!\om_0)^2\Delta Q_h\nonumber\\
        &=\;\om_0\, \om_1^3\, \big(2\!-\!u\!-\!v \big )\Big[2-\tcs-\ths\Big]\nonumber\\
        &+\om_0 \, \om_1^2 \,\Big[\om_0\ths+(2\om_1\!-\!\om_0)\tcs\Big] \nonumber\\
        &\;\;\times\big( u(1\!-\!u)+v(1\!-\!v)\big).
    \end{align}
 Noticing that $u,v\in [0,1]$ and $\om_1\!>\!\om_0$, regardless of the operating regime, we immediately conclude that both the terms on the right hand side of the above equation are individually non-negative and hence we obtain the inequality $\om_1^2\Delta W\!-\!(\om_1\!-\!\om_0)^2\Delta Q_h \geq 0$. The equality occurs in the quasistatic limit, \emph{i.e.}, for $u\!=\!v\!=\!1$. Repeating the same approach for the fluctuations of the heat from the cold reservoir we find that 
 \begin{align}
	&\om_1^2\Delta W-(\om_1\!-\!\om_0)^2\Delta Q_c\nonumber\\
	&=\l\frac{(\om_1\!-\!\om_0)^3}{2}(\om_1\!+\!\om_0)+\om_0\,\om_1^3 \, (2\!-\!u\!-\!v)\r\nonumber\\
	&\quad\times\Big[2-\tcs-\ths\Big]\nonumber\\
	&+\big( u(1\!-\!u)+v(1\!-\!v)\big)\Big[\om_1^4\tcs\nonumber\\
	&\quad+\om_0^3\,(2\om_1\!-\!\om_0)\ths\Big]> 0,
\end{align}
where, once again, both terms on the right hand side of the above equation are individually always greater than zero  as long as $u,v\in [0,1]$ and $\om_1\!>\!\om_0$. Thus, as the first central result, we see that the ratios of non-equilibrium fluctuations are not arbitrary but constrained by the parameters of the Otto-cycle.  {Also, notice that these constraints hold true in all four operational regimes of the Otto-cycle. }

{In what follows, we will show that once we specify that the Otto-cycle operates in the engine or refrigerator regime, the fluctuations of heat and work receive additional upper bounds.} In the absence of underlying time-reversal symmetry ($u\neq v)$, since the forward and {reverse} cycles are no longer identical, we find that the consistent way of characterizing the engine (or refrigerator) operational regime requires imposing engine (or refrigeration) constraints on both time-forward and {reverse} cycles. In other words, we characterize the engine (or refrigerator) operating regime of the Otto cycle demanding that both time-forward and {reverse} cycles function as an engine (or refrigerator). Because of this specification, the bounds found in our work matches with the earlier bounds found in the time-symmetric case \cite{Saryal2021} in a certain limit ($u\!=\!v$). We emphasise that, the proper symmetrized definition of the fluctuations introduced in Eq.~\eqref{def-fluc} and \eqref{def-re} are crucial to demonstrate useful bounds.

Before we proceed, let us briefly summarize the conditions for the quantum Otto cycle to function as an engine or a refrigerator.  Given that $\beta_c>\beta_h$ and $\om_1>\om_0$, the required and sufficient conditions for the Otto-cycle to function as an engine {in both time-forward and reverse cycles} are $\la W\ra_{F}\le0$ and $\la W\ra_{R}\le0$. These criteria impose constraints on the values of  $u$ and $v$, which together with the first law of thermodynamics determine the signs of the average heat flows, $\la Q_h\ra_{F(R)}\ge0$ and $\la Q_c\ra_{F(R)}\le0$. In the quasistatic limit ($u\!=\!v\!=\!1$), the engine condition simplifies to $\beta_c \om_0\ge\beta_h\om_1$. On the other hand, the necessary and sufficient conditions to realize refrigeration are $\la Q_c\ra_{F}\ge0$ and $\la Q_c\ra_{R}\ge0$. These, combined with the first law immediately set, $\la W\ra_{F(R)}\ge0$ and $\la Q_h\ra_{F(R)}\le0$. {In the rest of this paper, the engine (or refrigerator) operation regime will imply that both the time-forward and reverse cycles operate as engine (or refrigerator).} 



\subsection{Bounds on fluctuation in the engine regime}
In a prior study \cite{Saryal2021}, it was demonstrated that the ratio of output work fluctuation to input heat fluctuation for a symmetrically driven qubit-Otto cycle operating as an engine is lower bounded by the square of the average efficiency, \emph{i.e.}, 
$\eta^{(2)}\!=\! \llangle W^2\rrangle/\llangle  Q_h^2\rrangle \ge\la\eta\ra^2$, for a  { symmetrically} driven ($u\!=\!v$) qubit-Otto cycle,
where, $\langle\eta\rangle\!=\!-\la W\ra/\la Q_h\ra$. The immediate effect of this finding is that the relative fluctuation of output work is consistently greater than that of the input heat. For time-asymmetric work protocols ($u \neq v$), we redefine the quantities $\eta^{(2)}$ and average efficiency $\la\eta\ra$ by taking into account both the time-forward and {reverse} cycles
 \begin{align}
     \langle \eta\rangle&\coloneqq-\frac{\langle W\rangle_F+\langle W\rangle_R}{\langle Q_h\rangle_F+\langle Q_h\rangle_R}\\
     \eta^{(2)}&\coloneqq\frac{\Delta W}{\Delta Q_h}.
 \end{align}
 We discovered that $\eta^{(2)}$ is bounded from both above and below in the engine operational regime. 
\begin{align}\label{1central}
    1> \eta^{(2)}\ge \eta_{\text{otto}}^2\ge\langle \eta\rangle^2.
\end{align}
Eq.~(\ref{1central}) is the second central result of this work. Note that, the lower bound $\eta^{(2)}\ge \eta_{\text{otto}}^2$  follows simply from Eq.~(\ref{lbound}) (as the analysis is true in any operational regime), where,  $\eta_{\text{otto}}\!=\!1\!-\!\omega_0/\omega_1$, is the maximum attainable efficiency of the Otto engine considered here.  Importantly,  $\eta_{\text{otto}}^2$ sets a tighter lower bound $\eta^{(2)}$ in comparison to what was reported in \cite{Saryal2021}, and its value is fixed once we fix the parameter values to execute the engine-cycle. In the quasistatic limit the lower bounds on $\eta^{(2)}$ saturate, \emph{ i.e.}, $\eta^{(2)}\!=\!\eta_{\mathrm{otto}}^2\!=\!\la\eta\ra^2$. A further fundamental relationship between the {symmetrized} fluctuations is also provided by the upper bound on $\eta^{(2)}$: the {symmetrized} fluctuation of input heat is always greater than the fluctuation of output work, $\Delta Q_h>\Delta W $. Note that this upper bound also holds true for { time-symmetric} driving, where $u\!=\!v$. 
 Interestingly, for the time-asymmetric driving example, $u\neq v$,  we find violation of this upper bound if we only take into account the forward (or {reverse}) cycle, \emph{i.e.},  $\llangle W\rrangle_{F/R}/\llangle Q_h\rrangle_{F/R}\nleq 1$. 

 We now provide a rigorous proof for the upper bound $1>\eta^{(2)}$. We proceed by considering a modified version  of this inequality and prove that  $\Delta Q_h\!-\!\Delta W >0 $. The left hand side of this inequality is written as
\begin{align}
    &\Delta Q_h\!-\!\Delta W\nonumber\\
&= \frac{\omega_0}{2}\Big[2\!-\!\tcs\!-\! \ths\Big]{\cal A}+{\cal B}_1\!-\!{\cal B}_2,\label{qubit-upbound}
\end{align}
where, we have defined
\begin{align}
    {\cal A}&=-\big(\om_0+(1\!-\!2u)\om_1\big)\!-\!\big(\om_0+(1\!-\!2v)\om_1\big),\\
    {\cal B}_1&=\frac{\om_0^2}{2}\Big[2\!-\!\tcs\!-\!\ths\Big],\\
    {\cal B}_2&=\omega_0^2\big[u(1\!-\!u)+v(1\!-\!v)\big]\ths.
\end{align}
In order to show that $\Delta Q_h\!-\!\Delta W$ is always positive in the engine operational regime, we will use the generalised engine criteria (set up works as engine in both the forward and {reverse} cycles) to demonstrate that ${\cal A}>0$ and ${\cal B}_1-{\cal B}_2\ge0$. 

First, we observe that $\mathrm{min}(1\!-\!2u),\mathrm{min}(1\!-\!2v)\!=\!-\!1$, since $u,v\in[0,1]$. Therefore, from Eq.~(\ref{qubit-work}) we conclude that regardless of the operation regime, both $\langle W_3\rangle_F$ and $\la W_3\ra_R$ are always positive, since $\om_1>\om_0$. Now, once we fix that the set-up functions as an engine in both forward and {reverse} cycles, in order to satisfy the conditions on total work $\langle W\rangle_F\le0$ and $\langle W\rangle_R\le0$ we must have $\langle W_1\rangle_F,\langle W_1\rangle_R<0$, which immediately imply that $\big(\om_0+(1\!-\!2u)\om_1\big)<0$ and $\big(\om_0+(1\!-\!2v)\om_1\big)<0$. Thus, we conclude that quantity ${\cal A}$ is always positive. 

\begin{figure}[H]
    \centering
     \includegraphics[width=1.0\columnwidth]{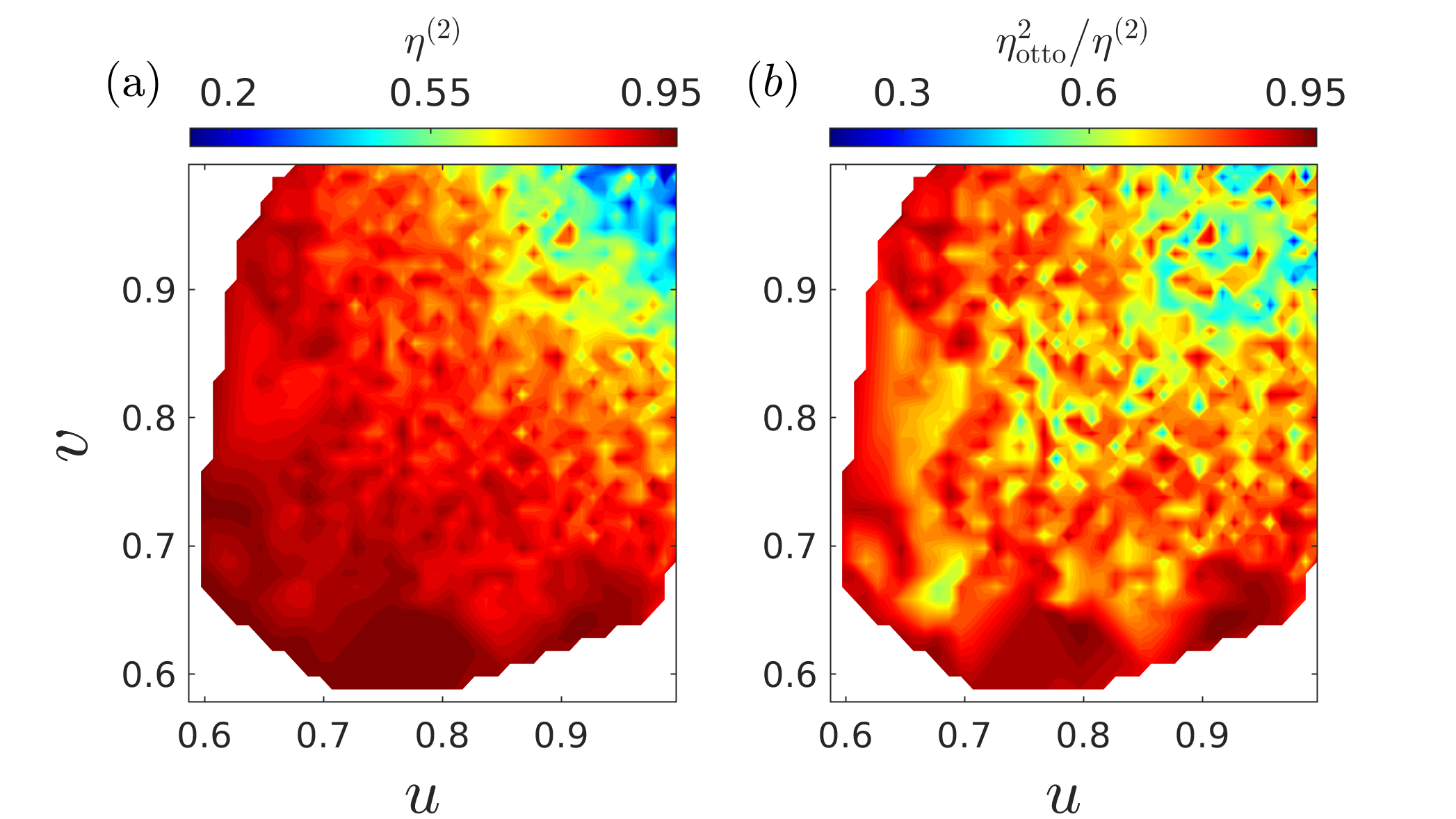}
     \caption{Results for (a) upper and (b) lower bounds on $\eta^{(2)}$ for an asymmetrically driven qubit-Otto cycle. $\om_0$,  $\om_1$, $T_c$ and $ T_h$ were chosen randomly from uniform distribution between the interval [0,5]. $u,v$ were chosen between [0,1]. Simulations done over 2.5 million points. {The white color regions correspond to situations where the generalized  engine conditions \emph{i.e.}, $\la W\ra_F\le0$ and $\la W\ra_R\le0$ are not satisfied.}}
     \label{fig:2}
\end{figure}

Next, to show that ${\cal B}_1\!-\!{\cal B}_2\ge0$, we first maximize  ${\cal B}_2$ with respect to the engine constraints $\langle Q_h\rangle_F\ge0$ and $\langle Q_h\rangle_R\ge0$. Notice that, the lower bounds of $u$ and $v$ are subjected to stricter restrictions under these constraints, and can be expressed as 
\begin{align}
    \frac{1}{2}+\frac{\th}{2\;\tc}\le u,v\le 1.
\end{align}
These lower limits in turn set an upper bound on the quantity $u(1\!-\!u)+v(1\!-\!v)$ denoted by the below inequality
\begin{align}
    u(1\!-\!u)+v(1\!-\!v)&\le \frac{1}{2}\frac{\tanh^2{\Big(\frac{\beta_c\omega_0}{2}\Big)}\!-\!\tanh^2{\Big(\frac{\beta_h\omega_1}{2}\Big)}}{\tanh^2{\Big(\frac{\beta_c\omega_0}{2}\Big)}}\nonumber\\
    &\le\frac{1}{2}\frac{\tanh^2{\Big(\frac{\beta_c\omega_0}{2}\Big)}-\tanh^2{\Big(\frac{\beta_h\omega_1}{2}\Big)}}{\tanh^2{\Big(\frac{\beta_h\omega_1}{2}\Big)}}.
\end{align}
Using the right hand side of the above inequality to maximize ${\cal B}_2$, we conclude that
\begin{align}
    {\cal B}_1\!-\!{\cal B}_2\ge{\cal B}_1\!-\!\big({\cal B}_2\big)_\mathrm{max}= \omega_0^2\Big[1\!-\!\tanh^2{\Big(\frac{\beta_c\omega_0}{2}\Big)}\Big]\ge0.
\end{align}
Now that we have shown that ${\cal A}>0$ and ${\cal B}_1\!-\!{\cal B}_2\ge0$, we immediately  infer  $\Delta Q_h\!-\!\Delta W> 0$. This concludes the proof of our second central result: Eq.~(\ref{1central}).   In Fig.~\ref{fig:2}, we provide a contour plot for the upper and lower bounds for $\eta^{(2)}$ by choosing arbitrary values for the various parameters drawn randomly from uniform  distributions. Both the bounds are respected for the asymmetrically driven case and thus match with our theoretical predictions.
\subsection{Proof for hierarchy of relative fluctuations in the engine regime}
Now that  Eq.~(\ref{1central}) has been demonstrated, the relation between the {symmetrized} relative fluctuations of total work and input heat (from the hot bath), is given as $\mathrm{RF}(W)\ge\mathrm{RF}(Q_h)$ which is readily obtained from the bound $\eta^{(2)} \geq \la \eta \ra^2$. This relation is valid in the engine {operational} regime. We now show that there exists a rigorous hierarchy between the {symmetrized} relative fluctuations of work and heat from both hot and cold reservoirs, in the engine operational regime, once the relative fluctuation of the exhaust heat into cold reservoir is taken into account. This hierarchy is stated as 
\begin{align}
    \mathrm{RF}(W)\ge\mathrm{RF}(Q_h)\ge\mathrm{RF}(Q_c).
    \label{hie-RF}
\end{align}
As the proof for $\mathrm{RF}(Q_h)\ge\mathrm{RF}(Q_c)$, we will demonstrate   the  modified  inequality given as
\begin{align}
    \frac{\big(\langle Q_c\rangle_F+\langle Q_c\rangle_R\big)^2}{\big(\langle Q_h\rangle_F+\langle Q_h\rangle_R\big)^2}\ge\frac{\Delta Q_c}{\Delta Q_h}.
\end{align}
As the first step, following the relation $\langle\eta\rangle\le\eta_{\text{otto}}$ together with the engine constraints $\langle Q_c\rangle_F,\langle Q_c\rangle_R\le0$, we immediately obtain, 
\begin{align}
\frac{\big(\langle Q_c\rangle_F+\langle Q_c\rangle_R\big)^2}{\big(\langle Q_h\rangle_F+\langle Q_h\rangle_R\big)^2}\ge \frac{\omega_0^2}{\omega_1^2}.\label{qh-qc-1}
\end{align}
Next,  in the engine operational regime we  show that,
\begin{align}
\frac{\omega_0^2}{\omega_1^2}\ge\frac{\Delta Q_c}{\Delta Q_h}.\label{qh-qc-2}
\end{align}
The proof for the modified version the above inequality,   $\omega_0^2\Delta Q_h\!-\!\omega_1^2\Delta Q_c\ge0$, goes as follows 
    \begin{align}
        &\omega_0^2\Delta Q_h-\omega_1^2\Delta Q_c\nonumber\\
        &=\big[u(1\!-\!u)+v(1\!-\!v)\big]\,\omega_0^2\,\omega_1^2\nonumber\\&\times\Big[\tanh^2{\Big(\frac{\beta_c\omega_0}{2}\Big)}\!-\!\tanh^2{\Big(\frac{\beta_h\omega_1}{2}\Big)}\Big]\ge0,
    \end{align}
as $\beta_c\om_0\ge\beta_h\om_1$ in the engine operational regime. Combining Eq.~(\ref{qh-qc-1}) and Eq.~(\ref{qh-qc-2}) we obtain
 \begin{align}
     \frac{\big(\langle Q_c\rangle_F+\langle Q_c\rangle_R\big)^2}{\big(\langle Q_h\rangle_F+\langle Q_h\rangle_R\big)^2}\ge \frac{\omega_0^2}{\omega_1^2}\ge\frac{\Delta Q_c}{\Delta Q_h}.
 \end{align}
 This concludes our proof for $\mathrm{RF}(Q_h)\ge\mathrm{RF}(Q_c)$. In the engine operational regime, this hierarchy between the relative fluctuations  as given in Eq.~(\ref{hie-RF}), is {{the third central result of our work}.  

Note that relative fluctuations of the integrated currents (average work and heat) can be linked to the precision of these quantities. The recently discovered TURs state that the precision of such non-equilibrium quantities is associated with a dissipation cost, which is quantified by the average entropy production \cite{ Potts_TUR, Horowitz_TUR_2019,Barato:2015:UncRel,trade-off-engine,Gingrich:2016:TUR,Falasco,Sacchi-TUR-HO}. In the following section we will first derive some important results concerning the TURs and make a further connection with the results obtained in the previous sections.
 
 \begin{figure}[H]
    \centering
     \includegraphics[width=1.0\columnwidth]{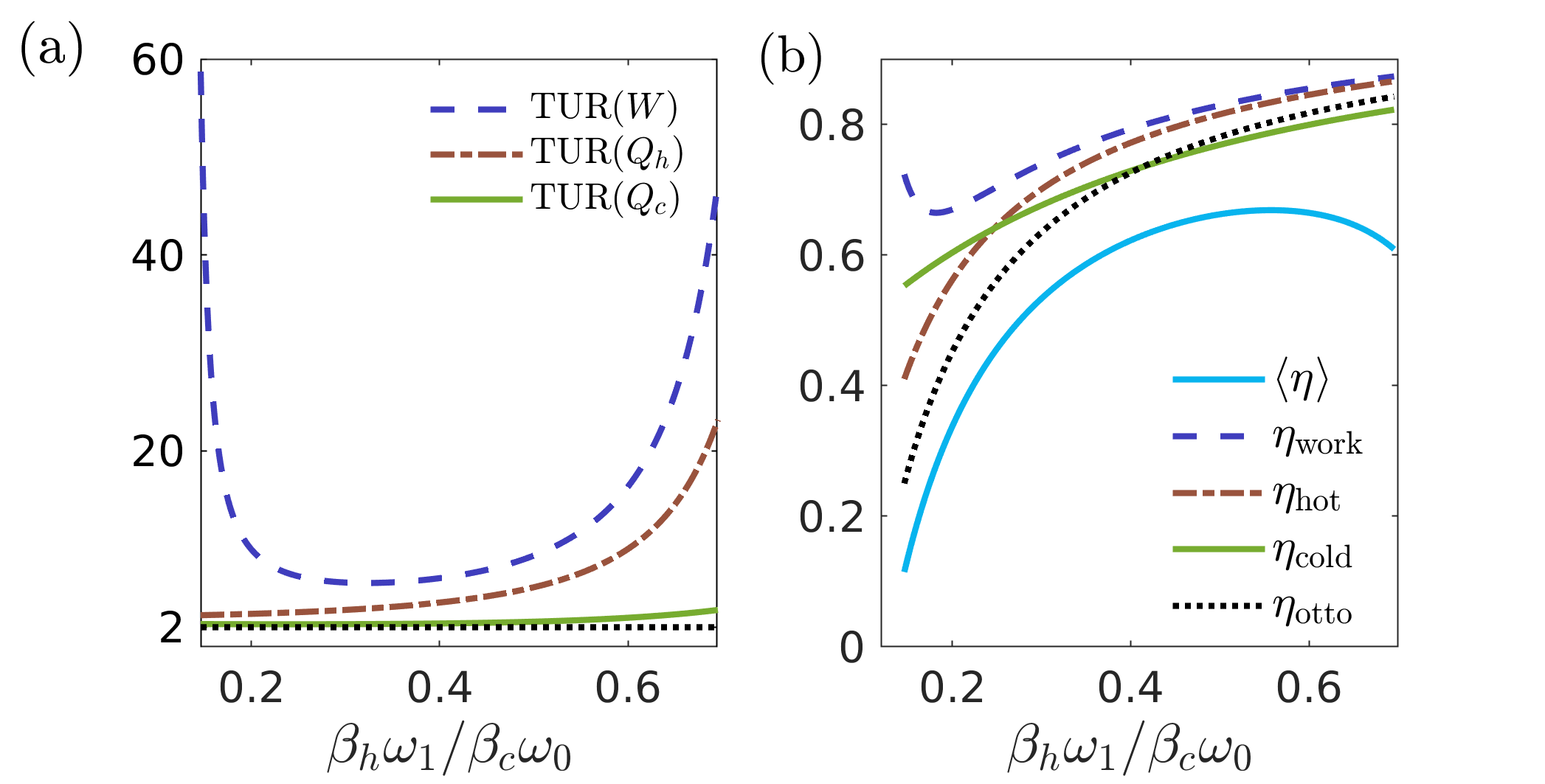}
     \caption{(a) Plot for the hierarchy between the three TURs in the engine regime. The black dashed line represents the lower bound 2. (b) Comparison between different bounds on average efficiency coming from the TURs. Parameters chosen: $T_c\!=\!0.57$,  $T_h\!=\!5.2$, $ \om_0\!=\!0.9$, $ u\!=\!0.99$ and $ v\!=\!0.9$. The black dashed line represents the ideal Otto efficiency $\eta_\mathrm{otto}$.}
     \label{fig:3}
\end{figure}

\subsection{Thermodynamic Uncertainty Relations for qubit-Otto cycle}
The thermodynamic uncertainty relations predict bounds on individual relative fluctuations in terms of the average entropy production $\langle \Sigma \rangle$. For the time-reversal symmetric case, a slightly modified version of TUR  was recently reported in Ref.~\cite{TUR-TLS} in the context of  a two-qubit swap engine. This modified TUR is given by
\begin{align}
    \frac{\llangle W^2\rrangle}{\la W\ra^2}\ge \frac{2}{\la \Sigma\ra}-1.\label{TUR-Sacchi}
\end{align}
In this work, we show analytically that the modified  TURs also hold for symmetrically driven four-stroke qubit-Otto cycle. Furthermore, we consider the general asymmetrically driven scenario, and demonstrate a proof for generalised TURs, given as
\begin{align}\label{TUR}
    \mathrm{TUR}(\phi)\equiv\frac{\la \Sigma\ra_F+\la \Sigma \ra_R}{2}\big(\mathrm{RF}(\phi)+1\big)\ge2.
\end{align}
with $\phi=W,\,Q_h$ and $Q_c$. As can be observed easily that these TURs in the time-reversal symmetric limit: $u\!=\!v$ take the form of Eq.~(\ref{TUR-Sacchi}). In Appendix \ref{sec:AppB}, we provide a rigorous proof of these generalised TURs  given in Eq.~(\ref{TUR}). Our proof is valid in arbitrary operational regime of the Otto-cycle, and for arbitrary asymmetric protocols. At this point, we emphasise that for this generic asymmetric driving situation, if we simply take time-forward (or {reverse}) cycle into account, TURs are violated. More specifically, numerical study revealed that, 
\begin{align}
    \la \Sigma\ra_{F/R}\l\frac{\llangle Q_c^2\rrangle_{F/R}}{\la Q_c\ra_{F/R}^2 }+1\r\ngeq2.
\end{align}
However, by accounting for both time-forward and {reverse} processes and constructing symmetrized expressions for the TURs, we can demonstrate that Eq.~(\ref{TUR})  always holds. 

Although, we save the rigorous proof of Eq.~(\ref{TUR}) for the Appendix \ref{sec:AppC} section, some key remarks are worth mentioning in the main text.

\noindent{\it Remark I} -- In the quasistatic limit: $u\!=\!v\!=\!1$, the  expressions of heat and work in the forward and {reverse} cycles become identical and we obtain
\begin{align}
\mathrm{TUR}(\phi)\big|_{u,v=1}=(\beta_c\omega_0\!-\!\beta_h\omega_1)\coth{\Big(\frac{\beta_c\omega_0\!-\!\beta_h\omega_1}{2}\Big)} \ge2\,,
\end{align}
where we have used the fact that $ x \coth(x) \geq 1$. Here $\phi=W,\,Q_h$ and $Q_c$ and  the $F,R$ subscripts have been removed. In the quasistatic limit, the TURs saturate as $\beta_c\om_0\!\to\!\beta_h\om_1$.

\noindent{\it Remark II} -- Away from the quasistatic limit, for general asymmetric driving scenario, although Eq.~(\ref{TUR}) still holds, the value of $\mathrm{TUR}(\phi)$ can become lesser than the quasistatic value depending on the parameters we choose to run the Otto-cycle.

Justifications of these statements are provided in Appendix \ref{sec:AppC}.
Interestingly, in the engine regime (for both forward and reverse cycle), where  we have already established the existence of a hierarchy between the {symmetrized} relative fluctuations of work and heat from both hot and cold reservoirs [see Eq.~(\ref{hie-RF})],  we can express all three TUR products in a nutshell--
\begin{align}\label{central3}
   \mathrm{TUR}(W)\ge\mathrm{TUR}(Q_h)\ge\mathrm{TUR}(Q_c)\ge2.
\end{align}
It is crucial to note that Eq.~(\ref{central3}) only applies to the engine operational regime.  {This is the fourth central result of this work.}} In Fig.~(\ref{fig:3})(a) we display the hierarchy in the generalised TURs in the engine regime with each TUR ratio  lower bounded by the value 2. In what follows, we show that, each of the TUR like trade-off relations offers thermodynamically consistent bound on {symmetrized} average efficiency $\la \eta\ra$ and we compare these bounds with the bound $\la \eta \ra \leq \eta_{\rm otto}$ as follows from Eq.~(\ref{1central}). 
\begin{figure}
    \centering
     \includegraphics[width=1.0\columnwidth]{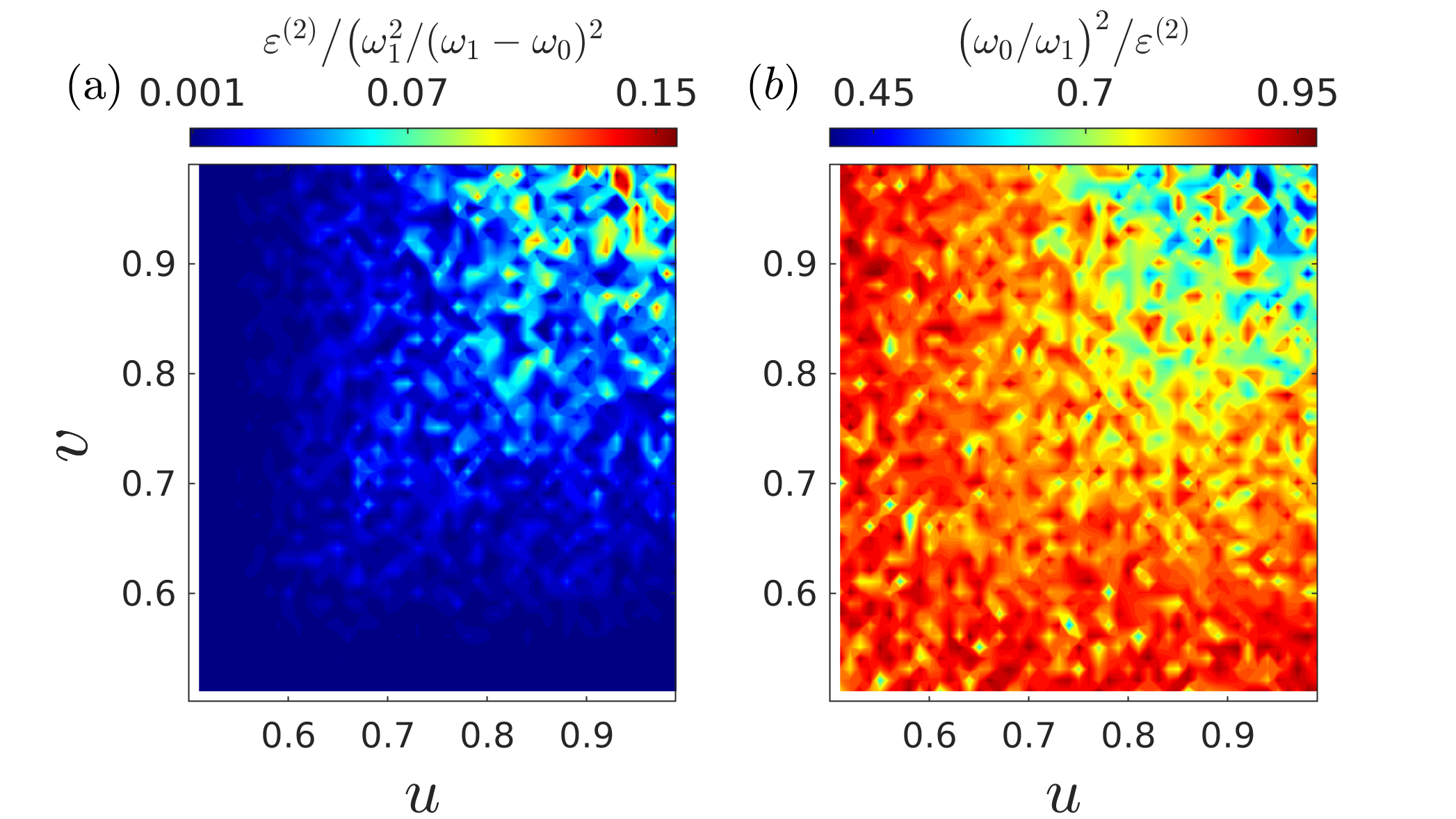}
     \caption{Results for (a) upper and (b) lower bounds on $\varepsilon^{(2)}$ in the refrigeration regime for asymmetrically driven qubit-Otto cycle. $\om_0$, $\om_1$, $T_c$ and $ T_h$ were chosen randomly from uniform distribution between the interval [0,5]. $u,v$ were chosen between [0,1]. Simulations done over 2.5 million points.}
     \label{fig:4}
\end{figure}
\subsection{Bounds on thermodynamic efficiency following the TURs}
In order to provide bounds on the average efficiency in the engine operational regime following the TURs, we first notice that the expression of symmetrized average entropy production can be re-written as,
\begin{align}
    \la \Sigma\ra_F\!+\!\la\Sigma\ra_R=&\;  \beta_c\big(\la Q_h\ra_F+\la Q_h\ra_R\big) \big(\eta_c\!-\!\la\eta\ra\big),\nonumber\\
   =& \!-\!\beta_c\big(\la W\ra_F+\la W\ra_R\big)\Big(1\!-\!\frac{\eta_c}{\la\eta\ra}\Big),\nonumber\\
    =&\!-\!\beta_c\big(\la Q_c\ra_F+\la Q_c\ra_R\big)\Big(1\!-\!\frac{1\!-\!\eta_c}{1\!-\!\la\eta\ra}\Big).\label{TUR-bound}
\end{align}
Now, by using the appropriate expression of entropy production from Eq.~(\ref{TUR-bound}), and plugging it into Eq.~(\ref{TUR}), we derived independent upper bounds on average efficiency for each $\phi=W,\,Q_h$ and $Q_c$.
We have listed them below: 
\begin{gather}
    \la \eta\ra\le \eta_{c}-2\frac{\la Q_h\ra_F+\la Q_h\ra_R}{\beta_c \, {A}(Q_h)}\equiv\eta_\mathrm{hot},\\
    \la \eta\ra\le {\eta_{c}}\Big/\l{1-2\frac{\la W\ra_F+\la W\ra_R }{\beta_c \, {A}(W)}}\r\equiv\eta_\mathrm{work},\\
    \la \eta\ra\le1-\frac{\beta_h}{\beta_c}\l1+2\frac{\la Q_c\ra_F+\la Q_c\ra_R}{\beta_c\, {A}(Q_c)}\r^{-1}\equiv\eta_\mathrm{cold}.\label{TURc-boundc}
\end{gather}
Here, $\eta_c\!=\!1\!-\!\beta_h/\beta_c$ is the Carnot efficiency and 
\begin{align*}
    {A}(\phi)=\Delta \phi+\frac{1}{2}\big(\la\phi\ra_F+\la\phi\ra_R\big)^2,
\end{align*}
with $\phi=W,\,Q_h$ and $Q_c$. Although the values of  $\mathrm{TUR}(\phi)$ in the engine regime follow a rigorous hierarchy as shown by Eq.~(\ref{central3}), the bounds on efficiency that arise from them disregard such hierarchical  relationship. However, numerical tests, as demonstrated in Fig.~(\ref{fig:3})(b), revealed that depending on the parameter regime, the bound $\eta_\mathrm{cold}$, derived from the tightest TUR, \emph{i.e.}, $\mathrm{TUR}(Q_c)\ge2$, can potentially become tighter than $\eta_\mathrm{otto}$ [see Eq.~(\ref{1central}) and Eq.~(\ref{TURc-boundc})].

\subsection{Bounds on fluctuation in the refrigerator regime}
We now briefly discuss the refrigerator operational regime. In case of refrigeration the central quantity to investigate is $\varepsilon^{(2)}\!=\! \Delta Q_c/\Delta W$. Eq.~(\ref{lboundc}) already provide an upper bound on $\varepsilon^{(2)}$. With refrigeration conditions, numerical results further suggest a lower bound on $\varepsilon^{(2)}$, and we write
\begin{equation}
    \frac{\om_1^2}{(\om_1\!-\!\om_0)^2}>\varepsilon^{(2)}>\frac{\om_0^2}{\om_1^2}.
\end{equation}
 In Fig.~\ref{fig:4}, we demonstrate the validity of the lower and upper bounds in the refrigeration regime of the qubit-Otto cycle. {Note that, unlike Eq.~(\ref{hie-RF}) which is valid in the engine regime}, we do not find hierarchy of symmetrized relative fluctuations in the refrigerator regime. This indicates fundamental differences in non-equilibrium fluctuations in different operational regimes and is a theme for future exploration. 
 
In what follows, we consider a parametrically driven harmonic oscillator as the working medium and illustrate the bounds.
\section{Harmonic oscillator under asymmetric driving as a working medium}
\label{sec:HO}
We consider a harmonic oscillator (HO), with unit mass, as the working fluid of the Otto-cycle operating between the two thermal reservoirs with inverse temperatures $\beta_c$ and $\beta_h$, with $\beta_c>\beta_h$. The parametric time-dependent Hamiltonian of the HO-working fluid is given by{
\begin{align}
    H[\om_t]=\frac{p^2}{2}+\frac{1}{2}\omega^2_t x^2
\end{align}
}For the cycle, during the unitary expansion stroke $A\!\to\! B$, the time-dependent trapping frequency $\omega_t$ goes from $\omega_0$ at $t\!=\!0$ to $\omega_1$ at time $t\!=\!\tau_1$, governed by the expansion protocol  $U_\mathrm{exp}$. During the compression stroke $C\!\to\! D$, the compression protocol  $U_\mathrm{com}$ drives $\omega_1$ back to $\omega_0$ in the time interval  $\tau_3$. The heat exchange strokes take place in between and perfect thermalization is achieved in both the heat exchange strokes. To assure time-asymmetric driving,   we again consider {$U_\mathrm{exp}^{\dagger}\ne \Theta U_\mathrm{com}\Theta^\dagger$.} For HO-Otto cycle, the exact joint CF of output work and input heat can be obtained and much like the qubit case, the non-adiabaticity of  asymmetric drivings  are captured by the parameters ${\cal Q}$ and ${\cal Q}^*$, where ${\cal Q},\,{\cal Q}^*\in[1,\infty]$ .Notice that, ${\cal Q}\!=\!{\cal Q}^*$ corresponds to the symmetric driving situation and ${\cal Q}\!=\!{\cal Q}^*\!=\!1$ is the quasi-static limit (see Appendix \ref{sec:AppD} for the details). Below we illustrate the numerical results. 

\begin{figure}[H]
    \centering
     \includegraphics[width=1.0\columnwidth]{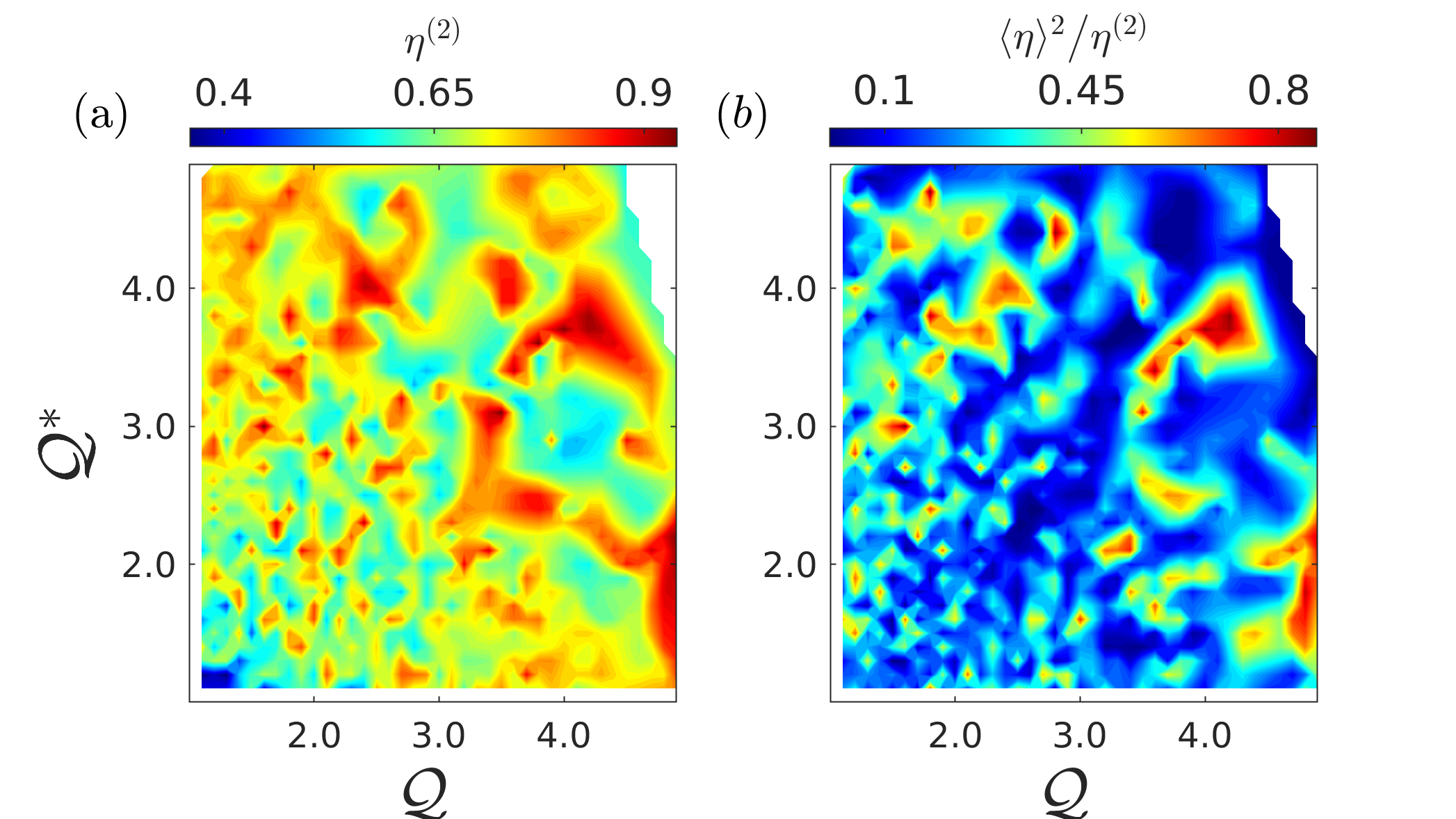}
     \caption{For an asymmetrically driven HO-Otto cycle (a) upper and (b) lower bounds on $\eta^{(2)}$ in the engine regime. $\om_0$, $\om_1$, $T_c$ and $ T_h$ were chosen randomly from uniform distribution between the interval [0,5]. ${\cal Q},{\cal Q}^*$ were chosen between [1,6]. Simulations done over 2.5 million points.}
     \label{fig:5}
\end{figure}

In the engine operational regime, $\la W\ra_F\le0$ and $\la W\ra_R\le0$, the numerical results in  Fig.~(\ref{fig:5}) suggest that the lower and upper bounds for $\eta^{(2)}$ are given as   $1>\eta^{(2)}\ge\la\eta\ra^2$. Contrary to Eq.~(\ref{1central}) for qubit-Otto cycle, numerical studies suggested that for the HO-Otto engine $\eta_\mathrm{otto}^2=\big(1\!-\!\om_0/\om_1\big)^2$ fails to provide a lower bound on $\eta^{(2)}$  at very low temperatures of the heat reservoirs, specifically for $T_c<1.0$. This discrepancy in results is caused by the different dimensionalities (number of energy levels) of the working fluids. Note that, $\eta_\mathrm{otto}$ corresponds to  the quantum-adiabatic (quasistatic) cycle's efficiency in both scenarios: $u\!=\!v\!=\!1$ for the  qubit-Otto cycle and ${\cal Q}\!=\!{\cal Q}^*\!=\!1$ for the HO-Otto cycle. However, as there are an infinite number of energy levels available for HO-working medium, quantum non-adiabaticity (transitions between the instantaneous eigen basis) plays a more significant role in determining the fluctuations and diminishes the usefulness of the adiabatic (quasistatic) value of efficiency $\eta_\mathrm{otto}$ in establishing bound on $\eta^{(2)}$.
\section{Summary}
\label{sec:summary}
In summary, we provide a study of bounds on non-equilibrium fluctuations of heat and work for asymmetrically driven four stroke quantum Otto engine with working fluid consisting of a qubit, or a harmonic oscillator. We show that the non-equilibrium fluctuations are not arbitrary but are restricted. In the engine regime, the ratio of non-equilibrium fluctuations of work and heat from hot reservoir $\eta^{(2)}$ receive both upper and lower bounds. For both the qubit and oscillator case, the upper bound for $\eta^{(2)}$ remains the same whereas the tighter lower bound $\eta_{\text{otto}}^2$,  found in qubit-Otto engine, is violated for the oscillator case. For the qubit-Otto cycle, we further make an important connection of our result with the TURs and observe that in the engine regime, the TURs of work and heat for both cold and hot reservoirs follow a strict hierarchy and further lower bounded by the value 2. These results  further indicate an interesting possibility to receive a tighter estimate for the thermodynamic efficiency.   While preliminary results  \cite{Shastri2022} indicate that some of these bounds may also be satisfied with HO and qubit-Otto engines with finite-time thermalization, it will be interesting to explore the universality of these bounds for more complex working fluids and for other class of finite-time engines.


\section*{ACKNOWLEDGMENTS}
SM and BKA thank Sushant Saryal for useful discussions. SM also thanks Mitesh Modasiya for helpful suggestions. SM acknowledges financial support from the CSIR, India (File number: 09/936(0273)/2019-EMR-I). MS acknowledges financial support through National Postdoctoral Fellowship (NPDF), SERB file
no. PDF/2020/000992. BPV acknowledges support from the Department of Science \& Technology Science and Engineering Research Board (India) Start-up Research Grant No. SRG/2019/001585. BKA acknowledges the MATRICS grant MTR/2020/000472 from SERB, Government of India. BKA also thanks the Shastri Indo-Canadian Institute for providing financial support for this research work in the form of a Shastri Institutional Collaborative Research Grant (SICRG). BKA would also like to acknowledge funding from National Mission on Interdisciplinary Cyber-Physical Systems (NM-ICPS) of the Department of Science and Technology, Govt. Of India through the I-HUB Quantum Technology Foundation, Pune INDIA.

\renewcommand{\theequation}{A\arabic{equation}}

\renewcommand{\thesection}{A\arabic{section}}
\setcounter{equation}{0}
\appendix
{
\section{Statistics of heat from the cold bath $Q_c$:}
\label{sec:AppA}
In this appendix, we provide a brief demonstration on how to calculate the cumulants of $Q_c$ from $P (W, Q_h)$. In case of perfect thermalization we will prove that the stochastic variable $Q_c$ and the variable $\tilde{Q_c}\!=\!-W\!-Q_h$ have one and the same statistics.  To determine the statistics of all the stochastic variables $W_1$, $Q_h$, $W_3$ and $Q_c$, we need to introduce `diagnostic’ projective quantum measurements after every stroke and the beginning of the cycle \cite{talkner18}. Let the successive energy measurement results respectively be denoted by $E_0$, $E_1$, $E_2$, $E_3$, and $E_{0'}$. Clearly these energies take values from the sets of energy eigenvalues $\{\epsilon_k[\lm_i]\}$ and $ \{\epsilon_j [\lm_f]\}$
of the initial Hamiltonian $H[\lm_i] =\sum_k \epsilon_k[\lm_i] \hat{P}_k[\lm_i]$ $(E_0, E_3, E_{0'})$ and final Hamiltonian $H[\lm_f] =\sum_j \epsilon_j[\lm_f] \hat{P}_j[\lm_f]$ (E1, E2), respectively. Here $\hat{P}_
{k(j)}[\lm_{i(f)}] = |k(j); \lm_{i(f)}\ra\la k(j); \lm_{i(f)}|$
 are eigen-projectors. With these measured energies we can determine the heat exchanged with the cold bath as
\begin{align}
    Q_c=E_{0'}-E_3,
\end{align}
where, notice that $E_{0'}$ denotes the measurement result
after the final thermalization with the cold bath. Now the
key question we seek to answer is  what is the relation between $Q_c$ and
the variable
\begin{align}
    \tilde{Q}_c=-(W_1+W_3+Q_h)=E_0\!-\!E_3. 
\end{align}
For perfect thermalization or even imperfect thermalization once the engine goes to a steady limit cycle, from energy conservation (First law of thermodynamics) we expect the following relation to hold: $\la\tilde{Q}_c\ra\!=\!\la Q_c\ra$. Extending this expectation,
we now seek to answer when/if a more stricter condition, 
\emph{i.e.},  $\la\tilde{Q}_c^r\ra\!=\!\la Q_c^r\ra$ for any positive integer $r$ holds. In this
case we can conclude that $Q_c$ and $\tilde{Q}_c$ have one and the same statistics. The conditional probability (forward process) of the energy measurement results $E_0 \!=\! \epsilon_n[\lm_i]$, $E_1\! =\! \epsilon_m[\lm_f]$, $E_2 \!=\! \epsilon_k[\lm_f]$, $E_3 \!=\! \epsilon_l[\lm_i]$ and $E_{0'} \!=\! \epsilon_{n'}[\lm_i]$ is given as
\begin{align}
    p_F(n',l,k,m,n)\!=\!{\cal T}_{l\to n'}^{\beta_c}{\cal T}^{II}_{k\to l}{\cal T}_{m\to k}^{\beta_h}{\cal T}^{I}_{n\to m}p_{\beta_c}(n),\label{joint-transition}
\end{align}
where $p_{\beta_c}(n)\!=\!\exp{(-\beta_c\epsilon_n[\lm_i])}/{\cal Z}_c[\lm_i]$ is just the Gibbs canonical probability. The transition probabilities
in the two work strokes were introduced in Eq.~\eqref{transition-probability} in the main text. Coming to the transition probabilities in the heat strokes, let us first consider the hot bath stroke and write the transition matrix for the general heat stroke (not necessarily perfect thermalization)
\begin{align}
    {\cal T}_{m\to k}^{\beta_h}=\trace{\{\hat{P}_k[\lm_f]\Phi_{\beta_h}\big[\hat{P}_m[\lm_f]\big] \}}\label{hot-general}
\end{align}
where $\Phi_{\beta_h}[.]$ is the CPTP map representing the thermalization, \emph{i.e.}, the map representing the time evolution generated by the Lindblad master equation for the system in contact with a thermal bath. Only when we have perfect thermalization the result of this map becomes independent of the (normalized) input state $\hat{P}_m[\lm_f]$ and always
outputs the thermal density matrix, \emph{i.e.}, for perfect thermalization we have $\Phi_{\beta_h}[\hat{\rho}]\!=\!\exp{(-\beta_h H[\lm_f])}/{\cal Z}_h[\lm_f]$ and we have
\begin{align}
    {\cal T}_{m\to k}^{\beta_h}=p_{\beta_h}(k)\label{hot-perfect},
\end{align}
where $p_{\beta_h}(k)\!=\!\exp{(-\beta_h\epsilon_k[\lm_f])}/{\cal Z}_h[\lm_f]$. In analogy, we can
write down the transition matrix for the cold bath heat stroke
in the general case
as
\begin{align}
    {\cal T}_{l\to n'}^{\beta_c}=\trace{\{\hat{P}_{n'}[\lm_i]\Phi_{\beta_c}\big[\hat{P}_l[\lm_i]\big] \}}\label{cold-general}.
\end{align}
For perfect thermalization this reduces to 
\begin{align}
    {\cal T}_{l\to n'}^{\beta_c}=p_{\beta_c}(n')\label{cold-perfect}.
\end{align}
With this, we can write the (marginal) probability distributions for the two variables of interest $Q_c$ and $\tilde{Q}_c$ as
\begin{align}
    P_F(Q_c)\!&=\!\sum_{n',l,k,m,n}\delta\big(Q_c\!-\!\epsilon_{n'}[\lm_i]\!+\!\epsilon_l[\lm_i]\big)p_F(n',l,k,m,n),\\
       P_F(\tilde{Q}_c)\!&=\!\sum_{n',l,k,m,n}\delta\big(\tilde{Q}_c\!-\!\epsilon_{n}[\lm_i]\!+\!\epsilon_l[\lm_i]\big)p_F(n',l,k,m,n).
\end{align}
Using the above distributions and the transition joint
probability in Eq.~\eqref{joint-transition}, we can write down the r-{\it th} (raw)
moments of the variables as
\begin{align}
    \la Q_c^r\ra_F\!=\!\sum_{n',l,k,m,n}\big(\epsilon_{n'}[\lm_i]\!-\!\epsilon_l[\lm_i]\big)^r{\cal T}_{l\to n'}^{\beta_c}&{\cal T}^{II}_{k\to l}{\cal T}_{m\to k}^{\beta_h}\nonumber\\
    &{\cal T}^{I}_{n\to m}p_{\beta_c}(n)\label{Qc-PD}\\
        \la \tilde{Q}_c^r\ra_F\!=\!\sum_{n',l,k,m,n}\big(\epsilon_{n}[\lm_i]\!-\!\epsilon_l[\lm_i]\big)^r{\cal T}_{l\to n'}^{\beta_c}&{\cal T}^{II}_{k\to l}{\cal T}_{m\to k}^{\beta_h}\nonumber\\
    &{\cal T}^{I}_{n\to m}p_{\beta_c}(n)\label{Qc'-PD}
\end{align}
From the above two equations, it immediately becomes
clear that for transition matrix with
imperfect thermalization, where the transition probability depends on initial state-- as given by Eqs.~\eqref{hot-general}, \eqref{cold-general},
 we obtain
\begin{align}
    \la Q_c^r\ra_F\neq\la\tilde{Q}_c^r\ra_F.
\end{align}
Let us now consider that the heat strokes lead to perfect thermalization, then we can use the expressions in Eqs.~\eqref{hot-perfect} and \eqref{cold-perfect} for the transition probabilities in Eqs.~\eqref{Qc-PD} and
\eqref{Qc'-PD} which simplify as follows
\begin{align}
    \la Q_c^r\ra_F\!=\!\sum_{n',l,k,m,n}\big(\epsilon_{n'}[\lm_i]\!-\!\epsilon_l[\lm_i]\big)^r p_{\beta_c}(n')&{\cal T}^{II}_{k\to l}p_{\beta_h}(k)\nonumber\\
    &{\cal T}^{I}_{n\to m}p_{\beta_c}(n),\\
        \la \tilde{Q}_c^r\ra_F\!=\!\sum_{n',l,k,m,n}\big(\epsilon_{n}[\lm_i]\!-\!\epsilon_l[\lm_i]\big)^r p_{\beta_c}(n')&{\cal T}^{II}_{k\to l}p_{\beta_h}(k)\nonumber\\
        &{\cal T}^{I}_{n\to m}p_{\beta_c}(n).
\end{align}
To simplify this further, we note the following simple
property:
\begin{align}
   \sum_m &{\cal T}^{I}_{n \to m}=\sum_m\big|\la m ;\lambda_f|U_\mathrm{exp}|n;\lambda_i\ra\big|^2\nonumber\\
   &=\sum_m \la n ;\lambda_i|U_\mathrm{exp}^\dagger|m;\lambda_f\ra\la m ;\lambda_f|U_\mathrm{exp}|n;\lambda_i\ra\nonumber\\
   &=\la n;\lm_i|U_\mathrm{exp}^\dagger(\tau_1)U_\mathrm{exp}(\tau_1)|n;\lm_i\ra=1.
\end{align}
In addition, we use the fact that $\sum_n p_{\beta_c}(n)\!=\!\sum_{n'}p_{\beta_c}(n')\!=\!1$ to write
\begin{gather}
    \la Q_c^r\ra_F\!=\!\sum_{n',l,k}\big(\epsilon_{n'}[\lm_i]\!-\!\epsilon_l[\lm_i]\big)^r {\cal T}^{II}_{k\to l}p_{\beta_h}(k)p_{\beta_c}(n'),\\
    \la \tilde{Q}_c^r\ra_F\!=\!\sum_{l,k,n}\big(\epsilon_{n}[\lm_i]\!-\!\epsilon_l[\lm_i]\big)^r {\cal T}^{II}_{k\to l}p_{\beta_h}(k)p_{\beta_c}(n),
\end{gather}
giving the clear result
\begin{align}
    \la Q_c^r\ra_F=\la\tilde{Q}_c^r\ra_F.
\end{align}
For the {reverse} process, using similar analogy for the perfect thermalization in the two heat exchange strokes, it can be shown that 
\begin{align}
    \la Q_c^r\ra_R=\la\tilde{Q}_c^r\ra_R.
\end{align}
This is indeed the formal justification for the result that four consecutive measurements during the quantum Otto are enough to get the statistics of all three energy exchanges, as mentioned in the main text.}

\renewcommand{\theequation}{B\arabic{equation}}
\setcounter{equation}{0} 

\section{Proof for the Fluctuation symmetry}
\label{sec:AppB}
In this appendix, we provide a proof of Eq.~(\ref{Fluctuation-relation}). The PD for the time-forward process is given in Eq.~(\ref{PD}). The PD for the corresponding {reverse} process is given by{
\begin{align}\label{rev-PD}
     P_R&(W,Q_h)\!=\!\!\sum_{n,m,k,l}\!\delta \Big(W\!-\!\epsilon_k[\lm_f]\!+\!\epsilon_l[\lm_i]\!-\!\epsilon_n[\lm_i]\!+\!\epsilon_m[\lm_f]  \Big)\nonumber\\
     &\delta\Big(Q_h\!-\!\epsilon_m[\lm_f]\!+\!\epsilon_k[\lm_f]\Big)\tilde{{\cal T}}^{I}_{l\to k}\tilde{{\cal T}}^{II}_{m\to n} 
              \frac{e^{-\beta_c \epsilon_l[\lm_i]}}{{\cal Z}_c[\lm_i]}
              \frac{e^{-\beta_h \epsilon_m[\lm_f]}}{{\cal Z}_h[\lm_f]}.
\end{align}  
Note that, in the reverse cycle the expansion and compression strokes are governed by { $\tilde{U}_\mathrm{exp}\!=\!\Theta U_\mathrm{com}^\dagger\Theta^\dagger$ and $\tilde{U}_\mathrm{com}\!=\!\Theta U_\mathrm{exp}^\dagger\Theta^\dagger$}, respectively. As a result, the transition probabilities for the time-forward and {reverse} processes follow the relations
\begin{align}
     \tilde{{\cal T}}^{I}_{l \to k}=&\big|{\la}k;\lm_f|\tilde{ U}_\mathrm{exp}|l;\lm_i\ra\big|^2\nonumber\\=&\big|{\la}k;\lm_f|\Theta U^\dagger_\mathrm{com}\Theta^\dagger|l;\lm_i\ra\big|^2\nonumber\\=&\big|{\la}k;\lm_f|{ U^\dagger_\mathrm{com}}|l;\lm_i\ra^*\big|^2=\big|{\la}l;\lm_i|U_\mathrm{com}|k;\lm_f\ra\big|^2\nonumber\\=& {\cal T}^{II}_{k \to l},\\
     \tilde{{\cal T}}^{II}_{m \to n}=&{\cal T}^{I}_{n \to m}.
\end{align}
 Now, using the property of delta function, $\delta(a-b)f(a)=\delta(a-b)f(b)$, Eq.~(\ref{rev-PD}) can be rewritten as 
\begin{align}
          P_R(W,Q_h)
          =&\sum_{n,m,k,l}\! \delta \Big(W\!+\!\epsilon_m[\lm_f]\!-\! \epsilon_n[\lm_i]\!+\!\epsilon_l[\lm_i]\!-\!\epsilon_k[\lm_f]\Big)\nonumber\\&\quad\delta\Big(Q_h\!+\!\epsilon_k[\lm_f]\!-\!\epsilon_m[\lm_f]\Big){\cal T}^{I}_{n\to m}{\cal T}^{II}_{k\to l}\nonumber 
              \\&\quad\frac{e^{-\beta_c \big(\epsilon_n[\lm_i]-W-Q_h\big)}}{{\cal Z}_c[\lm_i]}
              \frac{e^{-\beta_h \big(\epsilon_k[\lm_f]+Q_h\big)}}{{\cal Z}_h[\lm_f]}\nonumber\\
              =&\exp{\big(\Sigma\big)}P_F(-W,-Q_h),
\end{align}
}where, $\Sigma=\beta_c W+(\beta_c\!-\!\beta_h)Q_h$. With the simple change of variables $W\!\to\!-W$ and $Q_h\!\to\!-Q_h$ we recover $P_R(-W,-Q_h)=\exp{(-\!\Sigma)}P_F(W,Q_h)$, which is identical to Eq.~(\ref{Fluctuation-relation}).



\renewcommand{\theequation}{C\arabic{equation}}
\setcounter{equation}{0}
\section{Analytical proof of TUR for qubit-Otto cycle:}
\label{sec:AppC}
In this appendix, we provide a rigorous proof of Eq.~(\ref{TUR}), regardless of the operational regime. First, we will discuss the symmetric driving case ($u\!=\!v$). Later in this section we generalize the proof for asymmetric driving.  Notice that, in the symmetric driving scenario the time-forward and {reverse} processes become identical and Eq.~(\ref{TUR}) reduces to 
\begin{align}\label{TUR_sym2}
    \frac{\llangle\phi^2\rrangle}{\la \phi\ra^2}\ge\frac{2}{\la\Sigma\ra}-1,
\end{align}
where  $\phi\!=\!W,\, Q_h$ and $ Q_c$. Here we will consider an alternative expression
\begin{align}\label{TUR-sym}
  \la\Sigma\ra\big( \llangle \phi^2\rrangle+\la\phi\ra^2\big)\!-\!2\la \phi\ra^2
  \ge0
\end{align}
 and demonstrate the proof for $\phi\!=\!Q_c$. For $\phi\!=\!W$ and $Q_h$, the same analogy holds. 
In this particular section, for simplicity of notations, we define $x\!=\!\beta_c\om_0/2$ and $y\!=\!\beta_h\om_1/2$, and $t_{x}\!=\!\tanh(x)$, $t_y\!=\!\tanh(y)$.
\subsection{Symmetric driving case}
First, let us list the required expressions of first and second moments of $Q_c$ and the average entropy production using simplified notations
\begin{gather}
    \la \Sigma\ra=(x\!-\!y)(t_x\!-\!t_y)+2(1\!-\!u)(x t_y+y t_x),\\
    \la Q_c\ra = -\frac{\om_0}{2}\big(t_x\!-\!t_y\big)\!-\!\om_0\big(1\!-\!u\big)t_y,\\
    \la Q_c^2\ra=\llangle Q_c^2\rrangle+\la Q_c\ra^2=\frac{\om_0^2}{2}\big(1\!-\!t_x t_y\big)+\om_0^2(1\!-\!u)t_x t_y.
\end{gather}
Notice that, we have separated the quasistatic {(quantum-adiabatic)} part from the non-adiabatic contributions. With the above expressions in hand, the left hand side of Eq.~(\ref{TUR-sym}), apart from the positive factor $\om_0^2/2$, can be expressed as a quadratic function of the variable $p=2(1\!-\!u)$
\begin{align}
    f(p)=c+bp+ap^2.
\end{align}
Here
\begin{align}
  c=&(x\!-\!y)(t_x\!-\!t_y)(1\!-\!t_x t_y)\!-\!(t_x\!-\!t_y)^2,\\
  b=&t_x t_y (x\!-\!y)(t_x\!-\!t_y)\!+\!(1\!-\!t_x t_y)(x t_y\!+\!y t_x )\!-\!2t_y (t_x\!-\!t_y),\\
  a=&t_x t_y(x t_y\!+\!y t_x )\!-\!t_y^2,
\end{align}
and $p\in[0,2]$. Notice that, $p\!=\!0$ corresponds to the quasistatic situation. Our aim is to show that $f(p)$ is non-negative in the interval $[0,2]$. First, we check the signs of the coefficients.

{\it Sign of $c$} -- Notice that, $f(p\!=\!0)=c$ and it corresponds to the quasistatic situation. It is straightforward to show that
\begin{align}
    c=(t_x\!-\!t_y)^2\big[(x\!-\!y)\coth{(x\!-\!y)}-1\big]\ge0,
\end{align}
since $x\coth{x}\ge1$. Therefore, in the quasistatic situation the TUR for $Q_c$ in Eq.~(\ref{TUR_sym2}) always holds. 

{\it Sign of $b$} -- Let us now focus on sign of $b$. The first two terms in the expression of $b$ are always non-negative. For, $x\le y$ we can trivially get that $b\ge0$. However, determining the sign of $b$, for $x> y$, requires more involved analysis. We proceed as follows 
\begin{align}
    b\ge&(1\!-\!t_x t_y)(x t_y\!+\!y t_x )\!-\!2t_y (t_x\!-\!t_y)\nonumber\\
    =&(t_x\!-\!t_y)\big[\coth{(x\!-\!y)}(x t_y+y t_x)\!-\!2t_y\big]\nonumber\\
    =&(t_x\!-\!t_y)\big[\coth{(x\!-\!y)}(x t_y\!-\!y t_x)\!+\!2\big(\coth{(x\!-\!y)}y t_x\!-\!t_y\big)\big]\nonumber\\
    =&(t_x\!-\!t_y)t_x t_y\big[\coth{(x\!-\!y)}(x\coth{x}-y \coth{y})\nonumber\\
    &\quad +2\big(\coth{(x-y)}(y \coth{y})-\coth{x}\big)\big]\nonumber\\
    \ge&0.
\end{align}
Notice that, in the first line we discarded the first term of $b$, which is always positive, and for the last step, we use the fact that we are considering  $x>y$ situation, and make use of  the following identities
\begin{align*}
    (1)\quad & t_x\ge t_y,\\
    (2)\quad & x\coth{x}\ge y\coth{y}\ge 1,\\
    (3) \quad & \coth{(x\!-\!y)}\ge \coth{x}.
\end{align*}
Thus, we conclude that $b\ge0$, regardless of the operational regime.

Importantly, the coefficient $a$ does not have any definite sign and for $a\!=\!0$ we trivially get $f(p)\ge0$ in the allowed regime. Proceeding to the second step, we now perform a local minima/maxima analysis in order to find the minimum functional value of $f(p)$ in the allowed domain $p\in[0,2]$. The first and second derivatives of $f(p) $ with respect to $p$ are given in standard forms 
\begin{gather}
    f^\prime(p)=b+2 a p,\nonumber\\
    f^{\prime\prime}(p)=2a.
\end{gather}
The minima/maxima is given by the solution of $f^{\prime}(p)=0$ and reads as
\begin{align}
    p^*=-\frac{b}{2a}.
\end{align}
{\it Possibility I} -- If $a>0$, since we already have $b\ge0$, we find that $p^*\le0$, which lies outside the valid regime [0,2]. Note that, $a>0$ also implies $f^{\prime\prime}(p^*)>0$. Thus, we conclude that the function $f(p)$ will posses a minima for $a>0$, but the minima will not fall inside the valid range of $p\in[0,2]$. This is the central part of our argument--  no local minima of the function $f(p)$ exists for $p\in(0,2]$.

\noindent{\it Possibility II} -- If $a<0$, we have $p^*\ge0$ and $f^{\prime\prime}(p^*)<0$. This corresponds to maxima, and this maxima may or may not fall inside [0,2].

Interestingly, from both these possibilities we come to the same conclusion, stated as-- within this given range of $p\in[0,2]$ the function $f(p)$ will reach its minimum value at one of the end points. We already know $f(0)=c\ge0$. Now we find that
\begin{align}
    f(2)=(t_x\!+\!t_y)^2\big[(x\!+\!y)\coth{(x\!+\!y)}-1\big]\ge f(0)\ge0.
\end{align}
Thus for $p\in[0,2]$, we conclude that $f(p)\ge f(0)\ge0$. This is equivalent to
\begin{align}\label{TUR-sym1}
    \la \Sigma\ra\la Q_c^2\ra-2\la Q_c\ra^2\ge\big(\la \Sigma\ra\la Q_c^2\ra-2\la Q_c\ra^2\big)\big|_{u=1}\ge0.
\end{align}
Therefore, for the symmetric driving case, TUR given in Eq.~(\ref{TUR_sym2}) always holds. Similar analysis can also be carried out  for $Q_h$ and $W$.
This concludes the proof of TURs for the symmetric driving case. Now we move to the proof for more generalised situation-- the asymmetric driving scenario.

\subsection{Asymmetric driving case}
In this broken time-reversal symmetry situation, we will consider symmetrized form of Eq.~(\ref{TUR-sym}) by taking into account both the time-forward and {reverse} cycles, which is also an alternative version of Eq.~(\ref{TUR}),
\begin{align}\label{TUR-alternative}
    \frac{\la \Sigma\ra_F\!+\!\la \Sigma\ra_R}{2}&\Big[\frac{\llangle Q_c^2 \rrangle_F\!+\!\llangle Q_c^2\rrangle}{2}+\Big(\frac{\la Q_c\ra_F\!+\!\la Q_c\ra_R}{2}\Big)^2\Big]\nonumber\\
    &-2\Big(\frac{\la Q_c\ra_F\!+\!\la Q_c\ra_R}{2}\Big)^2\ge0.
\end{align}
For this poof our approach will be similar to the previously discussed symmetric driving case. Remember that now we are dealing with $u\!\neq\! v$ situation, and because of that the left hand side of Eq.~(\ref{TUR-alternative}) will be a function of both $u$ and $v$, and $u,v\in[0,1]$. However, here we will work with the variables $\tilde{p}=2\!-\!u\!-\!v$ and $q=u\!-\!v$, where $\tilde{p}\in[0,2]$ and $q\in[-1,1]$. This is equivalent to working with centre of mass and relative coordinates and $q$ indeed quantifies the relative  asymmetry introduced between the two unitary work strokes of the Otto cycle. This two variable function (apart from the positive factor $2\om_0^2$) is given by
\begin{align}
    \tilde{f}(\tilde{p},q)=c + b\tilde{p}+a\tilde{p}^2-g q^2-h \tilde{p} q^2,
\end{align}
and the allowed domain of the function $\tilde{f}(\tilde{p},q)$ is a square in the $\tilde{p}\,q$--plane with vertices at (0,0), (1,1), (2,0) and (1,-1).Our goal is to show that $\tilde{f}(\tilde{p},q)\ge0$ for the allowed values of  $\tilde{p}$ and $q$.
\begin{figure}[H]
    \centering
     \includegraphics[width=1.0\columnwidth]{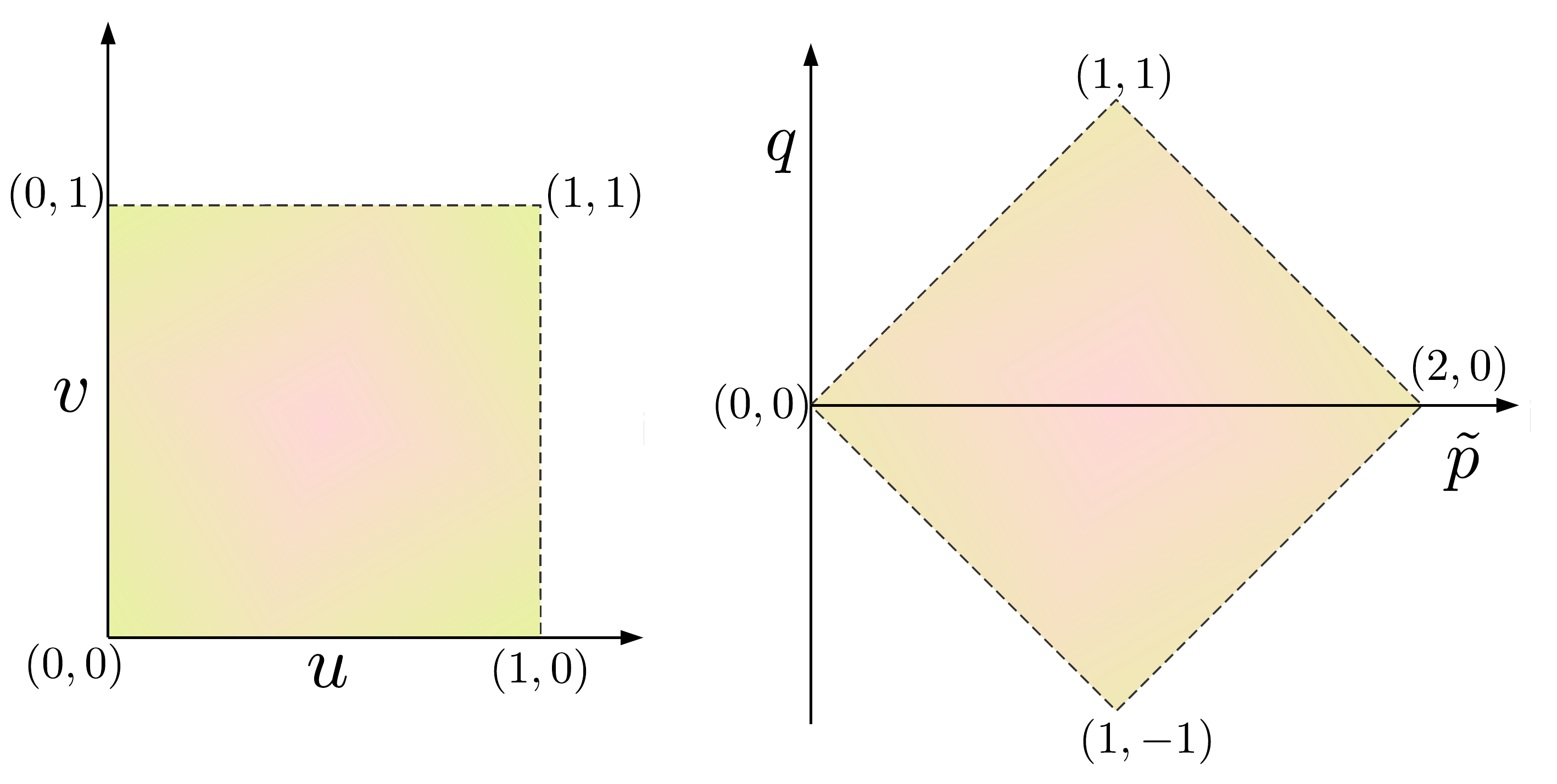}
     \caption{The allowed domain of the function $\tilde{f}(\tilde{p},q)$, where $\tilde{p}=2\!-\!u\!-\!v$ and $q=u\!-\!v$.}
     \label{fig:6}
\end{figure}
Notice that, we have already introduced the coefficients $c,b$ and $a$, and showed that $c,b\ge0$, whereas $a$ does not have definite sign. The new coefficients $g$ and $h $
are
\begin{align}
    g=&\frac{1}{2}(x\!-\!y)\,(t_x\!-\!t_y)\,t_y^2\,\ge0,\\
    h=&\frac{1}{2}(x t_y+y t_x)\,t_y^2\,\ge0.
\end{align}
 Note that, $\tilde{f}(0,0)\!=\!c$ corresponds to the quasistatic situation ($u\!=\!v\!=\!1$) and it is positive. We now proceed to the next step, \emph{i.e.}, find minimum possible value of $\tilde{f}(\tilde{p},q)$ for allowed $\tilde{p},q$ and show that the minimum value is non-negative. For this we perform maxima/minima analysis as before, but now we deal with two variables. Bellow we list the first and second partial derivatives of $\tilde{f}$ with respect to $\tilde{p}$ and $q$.
\begin{gather}
    \tilde{f}_{\tilde{p}}=b+2a\tilde{p}\!-\!h q^2,\nonumber\\
    \tilde{f}_q=-\!2q(g+h\tilde{p}),\nonumber\\
    \tilde{f}_{\tilde{p}\tilde{p}}=2a,\nonumber\\
    \tilde{f}_{q q}=-\!2(g+h\tilde{p}),\nonumber\\
    \tilde{f}_{\tilde{p} q}=-\!2h q.
\end{gather}
In order to find minima/maxima we solve $\tilde{f}_{\tilde{p}}\!=\!0$ and $\tilde{f}_q\!=\!0$ simultaneously. Now, the allowed solution of $\tilde{f}_q=0$ is $q^*\!=\!0$, as $g,h\ge0$ and $\tilde{p}$ cannot be negative. With this, solving $\tilde{f}_{\tilde{p}}=0$ provide us $\tilde{p}^*\!=\!-\!b/2a$. Now if $a>0$, we get $\tilde{p}^*\le0$, since $b\ge0$, and we immediately conclude that no maxima/minima or saddle point fall inside the allowed region in the $\tilde{p}q$--plane. However, if $a<0$, we have
\begin{gather}
    \tilde{f}_{\tilde{p}\tilde{p}}<0,\nonumber\\
    \tilde{f}_{\tilde{p}\tilde{p}}\tilde{f}_{q q}-\tilde{f}_{\tilde{p}q}^2>0,
\end{gather}
 and this corresponds to maxima. From  both cases we draw the same conclusion that, the function $\tilde{f}(\tilde{p},q)$ reaches its minimum allowed value at the boundary of its allowed domain.  So, now we examine the $\tilde{f}(\tilde{p},q)$ on boundary of the allowed square. Here, we will demonstrate calculations for one side $(0,0)\to(1,1)$ of the allowed square but, the conclusions hold true for the other three as well.  Notice that, this boundary line in the $\tilde{p}q$--plane is given by the equation $\tilde{p}=q$. Thus, on this boundary $\tilde{f}(\tilde{p},q)$ is parameterized as 
 \begin{align}
     \tilde{f}(\tilde{p})=c+b\tilde{p}+(a-g)\tilde{p}^2-h\tilde{p}^3.
 \end{align}
The derivatives of $\tilde{f}(\tilde{p})$ are given by
\begin{gather}
    \tilde{f}^\prime(\tilde{p})=b+2(a-g)\tilde{p}\!-\!3h\tilde{p}^2,\\
    \tilde{f}^{\prime\prime}(\tilde{p})=2(a-g)-6h\tilde{p}.
\end{gather}
Now, minima/maxima on this boundary is attained at $\tilde{f}^\prime(\tilde{p})=0$ which now quadratic in $\tilde{p}$. However,  since negative values of $\tilde{p}$ are not allowed, the only possible solution is
\begin{align}
    \tilde{p}^*=\frac{(a-g)+\sqrt{(a-g)^2+3 h b}}{3 h}.
\end{align}
Interestingly, for this value we find that 
\begin{align}
    \tilde{f}^{\prime\prime}(\tilde{p}^*)=-2\sqrt{(a-g)^2+3 h b}<0,
\end{align}
which corresponds to maxima. Thus we conclude that the function $\tilde{f}(\tilde{p})$ attains its minimum value at one of the end points of line, namely (0,0) or (1,1). Performing the same analysis for the other three sides we discover that, surprisingly the  whole minimum value finding problem of this two variable function eventually boils down to examining its values at the vertices of the allowed square on the $\tilde{p}\,q$--plane. It is straightforward to show that  
\begin{align}
    \tilde{f}(0,0)=&(t_x\!-\!t_y)^2\big[(x\!-\!y) \coth{(x\!-\!y)}\!-\!1\big]\ge0,\\
    \tilde{f}(2,0)=&(t_x\!+\!t_y)^2\big[(x\!+\!y) \coth{(x\!+\!y)}\!-\!1\big]\ge0,\\
    \tilde{f}(1,1)=&\tilde{f}(1,-1)=\frac{1}{2}(x t_x+y t_y)(2\!-\!t_y^2)\!-\!t_x^2\nonumber\\
    =&\frac{(2\!-\!t_y^2)}{2}\big(x t_x\!+\!y t_y\!-\!t_x^2\!-\!t_y^2\big)+\frac{t_y^2}{2}\big(2-t_x^2-t_y^2\big)\nonumber\\
    \ge&0,
\end{align}
where in the last line we have used the identity $x\ge\tanh{x}$ for $x>0$. Now, $\tilde{f}(2,0)\ge\tilde{f}(0,0)$, and remember that $\tilde{f}(0,0)$ represents the quasistatic situation. However, $\tilde{f}(1,1)$, although a positive quantity,  can potentially become lesser than $\tilde{f}(0,0)$ for certain parameter values. Because of this, unlike the symmetric driving case, here we do not 
receive relation like Eq.~(\ref{TUR-sym1}). It is worth mentioning that $\tilde{f}(1,1)$  is a very interesting scenario as it represents $u\!=\!1,v\!=\!0$ case, meaning, the expansion stroke is done infinitely slowly (quasistatic) whereas the the compression happens instantly (quench).

Finally, we have proved that the function $\tilde{f}(\tilde{p},q)$ attains its minimum value either at (0,0) or at (1,1) and (1,-1) and this minimum value is non-negative. This concludes our proof of Eq.~(\ref{TUR-alternative}).

\renewcommand{\theequation}{D\arabic{equation}}

\setcounter{equation}{0}
\section{Harmonic oscillator as a working medium:}
\label{sec:AppD}
In this appendix, we provide details about the  parametrically driven harmonic oscillator working medium executing a four-stroke Otto cycle. In the asymmetrically driven scenario,
the exact expression of the characteristic function of joint probability distribution of output work and input heat for the time-forward process is given by \cite{Deffner-thesis},
\begin{align}\label{CF_HO}
       \chi^{HO}_F&(\gamma_w,\gamma_h)=2(1\!-\!e^{-\beta_c\omega_0})(1\!-\!e^{-\beta_h\omega_1})\nonumber\\&\big[{\cal Q}(1\!-\!x_1^2)(1\!-\!y_1^2)+(1\!+\!x_1^2)(1\!+\!y_1^2)\!-\!4x_1y_1\big]^{-\frac{1}{2}}\nonumber\\
       &\big[{\cal Q}^*(1\!-\!x_2^2)(1\!-\!y_2^2)+(1\!+\!x_2^2)(1\!+\!y_2^2)\!-\!4x_2y_2\big]^{-\frac{1}{2}},
\end{align}
where,
\begin{eqnarray}
     x_1&=&\exp{\big[-\!\omega_0(\beta_c+i\gamma_w)\big]},\nonumber\\
     y_1&=&\exp{\big[i\omega_1(\gamma_w\!-\!\gamma_h)\big]},\nonumber\\
     x_2&=&\exp{\big[-\!\omega_1(\beta_h+i(\gamma_w\!-\!\gamma_h))\big]},\nonumber\\
     y_2&=&\exp{\big[i\omega_0\gamma_w\big]}\nonumber,
\end{eqnarray}
and ${\cal Q},{\cal Q}^*\in[1,\infty]$ are the so called adiabaticity parameters for the expansion and compression unitary strokes, respectively. In other words,  ${\cal Q}$ and ${\cal Q}^*$ serve as a qualitative indicator of the degree of non-adiabaticity (in the quantum sense) introduced into the unitary work strokes (see \cite{Husimi-HO} for details).
 ${\cal Q}\!=\!{\cal Q}^*$ corresponds to the time-symmetric driving case with  ${\cal Q}\!=\!{\cal Q}^*\!=\!1$ is the quasi-static limit.
In the time-asymmetric driving scenario (${\cal Q}\!\neq\!{\cal Q}^*$), the CF for the time-{reverse} cycle can be  calculated from Eq.~(\ref{CF_HO}), by interchanging ${\cal Q}$ and ${\cal Q}^*$. Bellow, we list the expressions of the averages and fluctuations of work and heat for the forward cycle.  
\begin{widetext}
\begin{gather}
     \langle Q_h\rangle_F=\frac{\omega_1}{2}\Big[\ch\!-\!{\cal Q}\cc \Big],\\
    \llangle Q_h^2\rrangle_F=-\frac{\omega_1^2}{4}\Big[2\!-\!\chs\!-\!(2{\cal Q}^2\!-\!1)\ccs\Big],\\
    \langle W \rangle_F =\langle W_1\rangle_F +\langle W_3\rangle_F=\frac{1}{2}\Big[({\cal Q}\omega_1\!-\!\omega_0)\cc+({\cal Q}^*\omega_0\!-\!\omega_1)\ch\Big] ,\\ 
    \llangle W^2\rrangle_F=\langle W_1\rangle_F^2 +\langle W_3\rangle_F^2\!-\!\frac{(\omega_0+\omega_1)^2}{2} +\frac{\omega_0\omega_1}{2}({\cal Q}+{\cal Q}^*)+\frac{\omega_1^2}{4}({\cal Q}^2\!-\!1)\ccs+\frac{\omega_0^2}{4}({{\cal Q}^*}^2\!-\!1)\chs,\\
    \langle Q_c\rangle_F=\frac{\omega_0}{2}\Big[\cc\!-\!{\cal Q}^*\ch \Big],\\
    \llangle Q_c^2\rrangle_F=-\frac{\omega_0^2}{4}\Big[2\!-\!\ccs\!-\!(2{{\cal Q}^*}^2\!-\!1)\chs\Big].
\end{gather}
\end{widetext}
For the {reverse} cycle, similar expressions for averages and fluctuations of work and heat can be obtained by swapping  ${\cal Q}$ and ${\cal Q}^*$. 

 \bibliographystyle{apsrev4-1}
 \bibliography{ottoref.bib}
\end{document}